\documentclass[12pt]{article}
\usepackage{fontenc}
\usepackage{graphicx}
\usepackage[cmex10]{amsmath}
\usepackage{enumerate}
\usepackage{tikz}
\usepackage{tkz-euclide}
\usetkzobj{all}
\usetikzlibrary{fpu}
\usetikzlibrary{calc}
\usepackage{amssymb}
\usepackage{amsthm}
\usepackage{MnSymbol}
\usepackage{subfig}
\usepackage{parskip}
\usepackage[sort]{natbib}
\usepackage{url}
\usepackage{bm}
\usepackage{booktabs}
\usepackage{algpseudocode, algorithm}
\newtheorem{theorem}{Theorem}[section]
\newtheorem{lemma}[theorem]{Lemma}
\newtheorem{corollary}[theorem]{Corollary}
\newtheorem{proposition}[theorem]{Proposition}
\usepackage{setspace}
\newcommand{\argmin}{\operatornamewithlimits{argmin}}

\def\clap#1{\hbox to 0pt{\hss#1\hss}}

\newcommand{\hX}{\hat{\mathbf{X}}}
\renewcommand{\Pr}{\mathbb{P}}

\newtheorem{assumption}{Assumption}
\theoremstyle{definition}
\newtheorem{definition}{Definition}
\newtheorem*{remark}{Remark}
\numberwithin{equation}{section}

\makeatletter
\def\thm@space@setup{%
  \thm@preskip=\parskip \thm@postskip=0pt
}
\makeatother

\newcommand{\blind}{0}

\begin{document}

\def\spacingset#1{\renewcommand{\baselinestretch}%
{#1}\small\normalsize} \spacingset{1}

%%%%%%%%%%%%%%%%%%%%%%%%%%%%%%%%%%%%%%%%%%%%%%%%%%%%%%%%%%%%%%%%%%%%%%%%%%%%%%

\bibliographystyle{plainnat}
\if0\blind
{
  \title{\bf A semiparametric two-sample hypothesis testing problem for random graphs}
  \author{Minh~Tang, Avanti Athreya, Daniel L. Sussman, \\ Vince
    Lyzinski, and Carey
    E. Priebe \thanks{
      Minh Tang, Avanti Athreya and Carey E. Priebe are with the Department of Applied
      Mathematics and Statistics, Johns Hopkins University, Baltimore, MD 21218. 
      Vince Lyzinski is with the Johns Hopkins University Human Language Technology 
      Center of Excellence, Baltimore, MD 21211. Daniel L. Sussman is with the 
      Department of Statistics, Harvard University, Cambridge, MA 02138. 
      This work was partially supported by the Johns Hopkins
      University Human Language Technology Center of Excellence and the
      XDATA program of the Defense Advanced Research Projects
      Agency (DARPA) administered through Air Force Research Laboratory
      contract FA8750-12-2-0303.}}
\maketitle
} \fi

\if1\blind
{
  \bigskip
  \bigskip
  \bigskip
  \begin{center}

    {\LARGE\bf A semiparametric two-sample hypothesis testing problem for random graphs}
\end{center}
  \medskip
} \fi

\bigskip
\begin{abstract}
  Two-sample hypothesis testing for random graphs arises naturally in
  neuroscience, social networks, and machine learning. In this paper,
  we consider a semiparametric problem of two-sample hypothesis
  testing for a class of latent position random graphs. We formulate a
  notion of consistency in this context and propose a valid test for
  the hypothesis that two finite-dimensional random dot product graphs
  on a common vertex set have the same generating latent positions or
  have generating latent positions that are scaled or diagonal
  transformations of one another. Our test statistic is a function of
  a spectral decomposition of the adjacency matrix for each graph and
  our test procedure is consistent across a broad range of
  alternatives.  We apply our test procedure to real biological data:
  in a test-retest data set of neural connectome graphs, we are able
  to distinguish between scans from different subjects; and in the
  {\em C.elegans} connectome, we are able to distinguish between
  chemical and electrical networks.  The latter example is a concrete
  demonstration that our test can have power even for small sample
  sizes.  We conclude by discussing the relationship between our test
  procedure and generalized likelihood ratio tests.
\end{abstract}

% \ifCLASSOPTIONpeerreview
% \begin{center} \bfseries EDICS Category: SSP-SSAN, SSP-DETC \end{center}
% \fi

\noindent%
{\it Keywords:}
random dot product graph, semiparametric graph inference, two-sample hypothesis testing
\vfill

\newpage
\spacingset{1.45} % DON'T change the spacing!
%\maketitle

\section{Introduction}
\label{sec:introduction}
%Two-sample hypothesis testing for specific population parameters is a
%cornerstone of classical statistical inference. Furthermore, there
%exists a burgeoning literature on techniques for inference on random
%graphs, including vertex clustering, vertex classification and
%estimation of latent positions for a single graph
%\cite{newman2006modularity,Bickel2009,Choi2010,%
%  Snijders1997Estimation,Lyz_perfect_13,rohe2011spectral,sussman12};
%and graph matching \cite{zaslavskiy09,conte04:_thirt,vogelstein:_fast}
%and anomaly detection across multiple graphs \cite{heng2014}. 

The development of a comprehensive machinery for two-sample hypothesis
testing for random graphs is of both theoretical and practical
importance, with applications in neuroscience, social networks, and
linguistics, to name but a few.  For instance, testing for similarity
across brain graphs is an area of active research at the intersection
of neuroscience and machine learning, and practitioners often use
classical parametric two-sample tests, such as edgewise $t$-tests on
correlations or Mantel tests, or permutation tests on subgraphs, as
approaches to graph comparison \citep{richiardi_machine_learning,
richiardi_2011,sporns_complex, zalesky_wholebrain}. 
Our goal in this work is to provide a clear setting for a particular two-sample
graph testing problem and to exhibit a valid, consistent, tractable
test statistic. Our results provide, to the best of our knowledge, the
first principled approach to semiparametric two-sample hypothesis
testing on graphs.

We focus on a test for the hypothesis that two random dot product
graphs on the same vertex set, with known vertex correspondence, have
the same generating latent position or have generating latent
positions that are scaled or diagonal transformations of one another.
This framework includes, as a special case, a test for whether two
stochastic blockmodels have the same or related block probability
matrices. We use a spectral decomposition of the adjacency matrix to
estimate the parameters for each random dot product graph, and our
test statistic is a function of an appropriate distance between these
estimates.

In the two-sample graph testing problem we address,
the parameter dimension grows as the sample size grows. This
problem is not precisely analogous to classical two-sample tests for,
say, the difference of two parameters belonging to some fixed
Euclidean space, in which an increase in data has no effect on the
dimension of the parameter. The problem is also not 
nonparametric, since we view our latent positions as fixed
and impose specific distributional requirements on the
data---that is, on the adjacency matrices. Indeed, we regard the
problem as semiparametric, and we adapt the traditional definition of
consistency to this setting. In particular, we have power increasing
to one for alternatives in which the difference between the two latent
positions grows with the sample size.

As one example of the utility of the test procedures we describe, we
consider the problem of matching connectome data from {\em
  Caeronabdhitis elegans (C.elegans)}, a hermaphrodite worm whose
wiring diagrams have been widely studied
\citep{varshney11:_struc,hall91:_caenor,white86:_caenor}.  There are a
total of 302 neurons in the {\em C. elegans} brain and there are two
different---but related---neuronal networks, characterized by the
chemical wiring (chemical synapses) and electrical wiring (gap
junctions), with known vertex alignment between the networks. It is of
biological relevance to determine the extent to which the two wiring
diagrams are similar. This question can be framed in the context of
two-sample testing, and we provide one approach to its resolution.

{\em C. elegans} is an instance of a pair of graphs with a comparatively small
but aligned vertex set, and our numerical results on this specific data indicate
that our test procedure provides good power, despite a sample size in the
hundreds. Our numerical analysis on other simulated data affirms more broadly
that our test has power against a wide class of alternatives for moderate sample
sizes. The analysis of much larger data is also a pressing practical problem,
and connectome data representing pairs of graphs with known vertex alignment can
be on the order of $10^7$ vertices and $10^{10}$ edges \citep{gray12:_magnet}.
The existence of such large data sets indicates that there are practical
problems in which our theoretical guarantees apply.

As a smaller-scale example, we consider the test-retest diffusion MRI
data from the Kennedy-Krieger Institute (KKI) \citep{landman_KKI}. The
raw data consist of pairs of neural images from 21 subjects.  These
scans can be converted into graphs at various scales: smaller-scale
graphs are formed by regarding certain brain region as vertices and
edges as connections between them (with fibers in the brain estimated
by deterministic tractography).  Larger-scale graphs (i.e., those with
much finer resolutions) are then obtained by choosing certain voxels
(those that survive a certain masking procedure during the creation of
the smaller graphs) as vertices and edges as single fibers between
them.  See \citet{gray12:_magnet, gray_migraine} for additional
information on the construction of these graphs. The resulting graphs
range in size from 200,000 to 700,000 vertices.  Even though the
graphs are not precisely aligned, any pair of them share a subset of
vertices (these subsets can differ from pair to pair). We can thus
conduct pairwise tests to determine the similarities between these
scans.  Implementing our test on such pairs, we find, in general, that
we correctly identify scans belonging to the same patient and
distinguish between those belonging to different patients.  En route,
we devise a bootstrapping procedure, particularly suited for large
graphs, for the estimation of critical values.

While it may appear that the requirement of known vertex
correspondence between the graphs is a stringent one, the {\em
  C. elegans} and connectome data are but two examples of a diverse
class of such paired graphs for which subsequent inference is
key. Other examples include the comparison of graphs in a time series,
such as email correspondence among a group over time, the comparison
of document networks in multiple languages, or the comparison of user
behavior on different social media platforms.

We conclude the paper with a brief discussion of the applicability of
other test statistics, including intuitively appealing tests based on
the spectral or Frobenius norm of the difference of adjacency
matrices, and a discussion of the connection between our test and
classical generalized likelihood ratio tests.  Our test statistic is a
ratio whose numerator is a distance between the estimated and true
latent positions and whose denominator is related to the estimated
standard error. As such, it is in the spirit of a Wald test. Although
we endeavor to describe the strengths and weakness of several
different test statistics, our aim is not to provide a comprehensive
analysis of possible tests. The specific hypotheses we
consider are indicative of the multitude of questions that can arise
in the larger context of two-sample hypothesis testing on random
graphs.

The contributions of this paper is as follows. We formulate the
problem of two-sample hypothesis testing for random graphs. We propose
simple test procedures based on the embedding of the adjacency
matrices. We devise simple bootstrapping procedures
to estimate critical values for these test statistics. 
We derive a new and improved bound (see
Theorem~\ref{thm:conc_Xhat_X}) for the difference between the
estimated latent positions obtained from the embedding and the
original latent positions. 

\subsection{Related Work}
Hypothesis testing on a single graph has a long history, especially
when compared to the multiple-graph setting. Problems of clustering
and community detection for a graph can be framed as classical
parametric hypothesis tests.  To touch on several recent results, we
note that in \citet{Arias-Castro_ER}, the authors translate the problem
of community detection into a test for determining whether a graph is
Erd\"{o}s-R{e}nyi or whether it has an unusually dense subgraph. In
\citet{Rukhin11}, the authors provide a power analysis of the maximum
degree and size invariants for a similar problem, and in
\citet{bickel13:_hypot} the authors formulate the problem of
determining the number of communities in a network as a hypothesis
testing problem involving the number of blocks in a stochastic
blockmodel. In contrast, we consider a two-sample problem in a more
general setting.

The random dot product graph model generalizes both the stochastic
blockmodel (SBM) and degree-corrected SBM. Our results do not
directly apply to general latent position models, such as those
considered in \citet{Hoff2002}.  Nevertheless, to the extent that
latent positions can be estimated accurately in these alternative
models---itself a topic of current investigation---a distance between
estimated latent positions for two graphs on the same vertex set could
be used to derive appropriate hypothesis testing procedures. If the
two graphs are not on the same vertex set, or if the vertex
correspondence is unknown, other issues arise.  Finding the vertex
correspondence when one exists, but is unknown, is the problem of
``graph matching'' and is notoriously difficult
\citep{conte04:_thirt}.  It is possible that graph matching tools can
be used as a first step to align the graphs before employing our test,
but we do not consider this here. An alternate approach to
comparing graphs on potentially different vertex sets and with differing
numbers of vertices is the subject of the paper of
\cite{tang14:_two:nonparametric}. There, the latent positions
for the random dot product graph are viewed as being i.i.d from some pair of
underlying distributions, say $F$ and $G$, and the graphs comparison
translates to the nonparametric test of equality of $F$ and
$G$. 

Finally, for the two-sample hypothesis test we consider, one can also
construct test statistics using other embedding methods, such as
spectral decompositions of normalized Laplacian matrices. To prove
results similar to Theorem~\ref{thm:identity} through
Theorem~\ref{thm:1} for the Laplacian-based test statistics, however,
requires substantial technical machinery and non-trivial adaptation or
generalization of the results in
\citet{qin2013dcsbm,rohe2011spectral,chaudhuri12:_spect}, among
others.  Hence, for simplicity, we focus here on embeddings of the
adjacency matrix.

\section{Setting}
\label{sec:setting}
We focus here on two-sample hypothesis testing for the latent position
vectors of a pair of {\em random dot product graphs} (RDPG)
\citep{young2007random} on the same vertex set with a known vertex
correspondence, i.e., a bijective map $\varphi$ from the vertex set of
one graph to the vertex set of the other graph.  We shall assume,
without loss of generality, that $\varphi$ is the identity map. As we
have already remarked, the assumption of known vertex correspondence
is satisfied in a number of real-world problems. Random dot product
graphs are a specific example of {\em latent position random graphs}
\citep{Hoff2002}, in which each vertex is associated with a latent
position and, conditioned on the latent positions, the edges are
independent Bernoulli random variables with the mean parameters given
by a symmetric {\em link} function of the pairwise latent positions.
The link function in a random dot product graph is simply the dot
product.

\subsection{Random Dot Product Graphs}
We begin with a number of necessary definitions and notational
conventions.  First, we define a random dot product graph on
$\mathbb{R}^d$ as follows.

\begin{definition}[Random Dot Product Graph (RDPG)] 
  Let $\chi^n_d$ be defined by
\begin{equation*}
\chi^n_d=\{{\bf U} \in \mathbb{R}^{n\times d}: {\bf UU}^T \in
[0,1]^{n\times n} \, \textrm{ and } \,\textrm{rank}({\bf U})=d\}
\end{equation*} 
and let ${\bf X}=[X_1 \mid \cdots \mid X_n]^{T} \in \chi^n_d$.  Suppose ${\bf A}$ is
a random adjacency matrix given by
\begin{equation*}
\Pr[{\bf A}|{\bf X}]=
\prod_{i<j}(X_i^TX_j)^{{\bf A}_{ij}}(1-X_i^{T}X_j)^{1-{\bf A}_{ij}}
\end{equation*}
Then we say that ${\bf A} \sim \mathrm{RDPG}(\mathbf{X})$ is the adjacency 
matrix of a {\em random dot product graph} with {\em latent position} {\bf X} 
{\em of rank } $d$.
\end{definition}
We define the matrix ${\bf P}=(p_{ij})$ of edge probabilities by ${\bf
  P}={\bf XX}^T$.  We will also write ${\bf A} \sim
\mathrm{Bernoulli}({\bf P})$ to represent that the existence of an
edge between any two vertices $i,j$, where $i>j$, is a Bernoulli
random variable with probability $p_{ij}$; edges are independent. We
emphasize that the graphs we consider are undirected and
loop-free.

Suppose we are given two adjacency matrices ${\bf A}_1$ and 
${\bf A}_2$ for a pair of random dot product graphs on the same vertex
set. 
%We assume throughout this paper that the graphs are
%independent. 
Our goal is to develop a consistent, at most
level-$\alpha$ test to determine whether or not the two generating
latent positions are equal, up to an orthogonal
transformation. Indeed, if $\mathcal{O}(d)$ represents the collection
of orthogonal matrices in $\mathbb{R}^{d \times d}$ and if ${\bf W}
\in \mathcal{O}(d)$,then ${\bf XWW}^T{\bf X}^T={\bf P}$, leading to
obvious non-identifiability. %We also consider the cases of
%whether the latent positions are scalar multiples of one another and
%finally whether they are related by a diagonal linear transformation.

\subsection{Hypothesis Testing}
Formally, we state the following two-sample testing problems for
random dot product graphs.  Let ${\bf X}_n, {\bf Y}_n \in \chi^n_d$
and define ${\bf P}_n={\bf X}_n{\bf X}_n^T$ and ${\bf Q}_n={\bf
  Y}_n{\bf Y}_n^T$.  Given ${\bf A}
\sim \mathrm{Bernoulli }({\bf P}_n)$ and ${\bf B} \sim
\mathrm{Bernoulli}({\bf Q}_n)$, we consider the following tests:
\begin{enumerate}[(a)]
\item {\em (Equality, up to an orthogonal transformation)}
\begin{align*} H^n_0 \colon {\bf X}_n \upVdash {\bf Y}_n \quad 
     \text{against} \quad H^n_a \colon {\bf X}_n \nupVdash {\bf Y}_n
  \end{align*}
  where $\upVdash$ denotes that there exists an orthogonal matrix
  ${\bf W} \in \mathbb{R}^{d \times d}$ such that ${\bf X}_n={\bf
    Y}_n{\bf W}$.
  
\item {\em (Scaling)} 
\begin{align*}
  H^n_0 \colon {\bf X}_n\upVdash c_n{\bf Y}_n \textrm{ for some } c_n >0 \quad
  \text{against} \quad H^n_a \colon {\bf X}_n \nupVdash c_n{\bf Y}_n
  \textrm{ for any } c_n >0
\end{align*} 

\item {\em (Diagonal transformation)}
\begin{align*}
  H^n_0 \colon {\bf X}_n \upVdash \mathbf{D}_n {\bf Y}_n \textrm{ for some
    diagonal } {\bf D}_n \quad
  \textrm{ against } \quad H^n_a \colon {\bf X}_n & \nupVdash \mathbf{D}_n {\bf Y}_n
  \textrm{ for any diagonal } {\bf D}_n %\textrm{ with bounded positive entries}.
\end{align*} 
\end{enumerate}

In fact, throughout this paper, we will consider a sequence of such
tests for $n \in \mathbb{N}$.  We stress that in our sequential
formulation of (a) -- (c), the latent positions
$\mathbf{X}_n,\mathbf{Y}_n$ need not be related to $\mathbf{X}_{n'},
\mathbf{Y}_{n'}$ for any $n' \not = n$. However, the size of the
adjacency matrices $\mathbf{A}$ and $\mathbf{B}$ is quadratic in $n$
and hence the larger $n$ is, the more accurate are our estimates of
$\mathbf{X}_{n}$ and $\mathbf{Y}_n$.

To contextualize our choice of hypotheses, consider the specific case
of the stochastic blockmodel \citep{Holland1983} and the related
degree-corrected stochastic blockmodel \citep{karrer2011stochastic}.
Recall that a stochastic block model on $K$ blocks with block
probability matrix ${\bf N}$ can be viewed as a random dot product
graph whose latent positions are a mixture of $K$ fixed vectors.  In
(a), we test whether two stochastic blockmodel graphs $G_1$ and $G_2$
with fixed block assignments have the same block probability matrices
${\bf N}_1={\bf N}_2$.  In (b), we test whether the block probability
matrix of one graph is a scalar multiple of the other; i.e. if ${\bf
  N}_1=c{\bf N}_2$.  Finally, in (c), we test whether two
degree-corrected stochastic blockmodels have the same block
probability matrices, but possibly different degree-correction
factors.

We describe the test procedures for the above hypothesis tests in
more details in the next section. The main idea is that given suitable
estimates $\hat{\mathbf{X}}_n$ and $\hat{\mathbf{Y}}_n$ of
$\mathbf{X}$ and $\mathbf{Y}$, the associated test statistic is
essentially a function of $\min_{\mathbf{W} \in \mathcal{O}_{d}}
\|\hat{\mathbf{X}}_n - \hat{\mathbf{Y}}_n \mathbf{W}\|$.

\subsection{Adjacency spectral embedding and related results}
We now describe the adjacency spectral embedding of \citet{sussman12},
which serves as our estimate for the latent positions $\mathbf{X}$ and
$\mathbf{Y}$.

\begin{definition} 
  The {\em adjacency spectral embedding} (ASE) of $\mathbf{A}$ into
  $\mathbb{R}^{d}$ is given by $\hat{{\bf X}}={\bf U}_{\mathbf{A}}
  {\bf S}_{\mathbf{A}}^{1/2}$ where
$$|{\bf A}|=[{\bf U}_{\mathbf{A}}|\tilde{{\bf U}}_{\mathbf{A}}][{\bf
  S}_{\mathbf{A}} \bigoplus \tilde{{\bf S}}_{\mathbf{A}}][{\bf
  U}_{\mathbf{A}}|\tilde{{\bf U}}_{\mathbf{A}}]$$ is the spectral
decomposition of $|\bf{A}| = (\bf{A}^{T} \bf{A})^{1/2}$ and
$\mathbf{S}_{\mathbf{A}}$ is the matrix of the $d$ largest eigenvalues
of $|\mathbf{A}|$ and $\mathbf{U}_{\mathbf{A}}$ is the matrix whose
columns are the corresponding eigenvectors.
\end{definition}

Let ${\bf X}$ and ${\bf Y}$ be two latent positions in $\mathbb{R}^{n
  \times d}$, and let ${\bf A}\sim \mathrm{Bernoulli } ({\bf P})$ with
$\mathbf{P} ={\bf XX}^T$ and ${\bf B} \sim \mathrm{Bernoulli }({\bf
  Q})$ with $\mathbf{Q} = {\bf YY}^T$ represent the associated
adjacency matrices of the random dot product graphs with ${\bf X}$ and
${\bf Y}$, respectively, as their latent positions. We observe that
${\bf X}$, ${\bf Y}$, and ${\bf A}$ and ${\bf B}$ all depend on $n$,
but for notational convenience we will suppress this dependence except
when imperative for communicating an asymptotic property. Let
$\hat{{\bf X}}$ and $\hat{{\bf Y}}$ denote the corresponding adjacency
spectral embeddings of ${\bf A}$ and ${\bf B}$, respectively.  We use
$\|\cdot \|_F$ to denote the Frobenius norm of a matrix and $\|\cdot \|$
to denote the spectral norm of a matrix or the Euclidean norm of a
vector, depending on the context. % , and $\|\cdot\|_{2 \rightarrow
%   \infty}$ to denote the $2 \rightarrow \infty$ operator norm (that
% is, the maximum Euclidean norm of the rows of a matrix). 
Also, we define for a matrix $\mathbf{M}$ with singular values $\sigma_1(\mathbf{M}) \geq
\sigma_2(\mathbf{M}) \geq \dots$, the
parameters $\delta(\mathbf{M})$, $\gamma_1(\mathbf{M})$, and
$\gamma_2(\mathbf{M})$ as follows \begin{gather*}
  \delta(\mathbf{M}) = \max\limits_{1 \leq i \leq n} \sum_{j=1}^n
  M_{ij}; \qquad
  \gamma_1(\mathbf{M}) =\min_{ i \leq d} 
\frac{\sigma_i(\mathbf{M}) -
  \sigma_{i+1}(\mathbf{M})}{\delta(\mathbf{M})}; \qquad
  \gamma_2(\mathbf{M}) = 
  \frac{\sigma_{d}(\mathbf{M}) - \sigma_{d+1}(\mathbf{M})}{\delta(\mathbf{M})}
\end{gather*}
The definitions of $\gamma_1$ and $\gamma_2$ depends
implicitly on a parameter $d \in \mathbb{N}$; in this work, $d$ is
always assumed known and usually corresponds to the embedding
dimension for some adjacency spectral embedding. For a
matrix $\mathbf{P} = \mathbf{X} \mathbf{X}^{\top}$ of rank $d$,
$\delta(\mathbf{P})$ is simply the maximum expected degree of a graph
$\mathbf{A} \sim \mathrm{Bernoulli}(\mathbf{P})$,
$\gamma_1(\mathbf{P})$ is the minimum gap between the $d$ largest
eigenvalues of $\mathbf{P}$, normalized by the maximum expected degree
and $\gamma_2(\mathbf{P})$ is just
$\sigma_{d}(\mathbf{P})/\delta(\mathbf{P})$. It is immediate that
$\gamma_1 \leq \gamma_2$.

Throughout this work, our results depend on certain conditions on the
gap between the eigenvalues of $\mathbf{P}_n$ and certain minimum
sparsity conditions on $\mathbf{P}_n$ as $n$ increases. We state these
conditions in Assumption~\ref{eigengap_assump} below.  These
conditions are motivated by established bounds from
\cite{oliveira2009concentration,athreya2013limit,lyzinski13:_perfec}
on the separation between $\mathbf{A}$ and $\mathbf{P}$ and the
accuracy of the adjacency spectral embedding in the estimation of the
true latent positions.  We consolidate these known bounds in the
appendix, but in particular they imply
  \begin{equation}
    \label{procrustes_oldbound1}
    \min\limits_{\mathbf{W} \in \mathcal{O}(d)}\|\hat{\mathbf{X}}_n {\mathbf W} - \mathbf{X}_n\|_{F} = O(d \sqrt{\log{n}})
  \end{equation}  
with high probability.
%. In light
%of Proposition \ref{Old_bounds}, we consolidate our eigengap and
%sparsity assumptions on the sequence $\mathbf{P}_n$ as follows:
\begin{assumption}
\label{eigengap_assump} 
We assume that there exists a fixed $d \in \mathbb{N}$ such that for
all $n$, $\mathbf{P}_n$ is of rank $d$ with $d$ distinct positive
eigenvalues.  Further, we assume that there exist constants
$\epsilon>0$, $c_0>0$ and $n_0(\epsilon, c) \in \mathbb{N}$ such that
for all $n \geq n_0$:
\begin{align}
\label{eq:eigengap_gamma}
\gamma_1(\mathbf{P}_n) &> c_0\\
\label{eq:eigengap_delta}
\delta(\mathbf{P}_n) &> (\log n)^{2+ \epsilon}
% \gamma_2\sqrt{\delta} & > \sqrt{2 \log (n/\eta)}\\
% \sqrt{\gamma_1 \delta}& > \frac{4}{2 \sqrt{d \log n}} \, \texttt{
% (see marginal note on these assumptions)}
\end{align}
\end{assumption}
Because the parameters $\delta(\mathbf{P})$, $\gamma_1(\mathbf{P})$
and $\gamma_2(\mathbf{P})$ depend on ${\bf P}$, they cannot be
computed from the adjacency matrices alone. Therefore, we use the
corresponding estimates of these quantities, namely
$\delta(\mathbf{A})$, $\gamma_1(\mathbf{A})$, and
$\gamma_2(\mathbf{A})$. Proposition~\ref{prop:gamma(A)} of the
appendix guarantees the consistency of these estimates, and they also
provide a mechanism by which to check whether the conditions in
Assumption~\ref{eigengap_assump} hold.

%, since $\delta({\bf A}_n)$ can be used as an estimate for $\delta({\bf P}_n)$ 
We note that a level-$\alpha$ test can easily be generated from Eq.
\eqref{procrustes_oldbound1} itself. However, in the present work, we
provide an improved bound for $\|\hat{\mathbf{X}}-\mathbf{XW}\|$ that
is given in Theorem~\ref{thm:conc_Xhat_X} below. This new bound
enables us to describe more precisely the class of alternatives over
which the proposed test procedure is consistent.  In particular
Eq.~\eqref{procrustes_oldbound1} requires that for consistency, the
difference between the latent positions ${\bf X}_n$ and ${\bf Y}_n$
diverge at a rate of $\omega(\sqrt{\log{n}})$ as
$n \rightarrow \infty$; Theorem~\ref{thm:conc_Xhat_X} simply requires
that this difference diverges, with no restriction on the rate of
divergence. However, we reiterate that based on
Theorem~\ref{thm:conc_Xhat_X}, as $n$ grows, the test statistic we
construct will not always distinguish between two latent positions
${\bf X}_n$ and ${\bf Y}_n$ that differ in a constant number of rows.

\begin{theorem}\label{thm:conc_Xhat_X}
  Suppose $\mathbf{P}=\mathbf{XX}^T$ is an $n \times n$ probability
  matrix of rank $d$ and its eigenvalues are distinct.  Suppose also
  that there exists $\epsilon>0$ such that
  $\delta(\mathbf{P})>(\log{n})^{2 + \epsilon}$. Let $c>0$ be
  arbitrary but fixed. Then there exists a $n_0(c)$ and a universal
  constant $C \geq 0$ such that if $n \geq n_0$ and $n^{-c} <
  \eta<1/2$, then there exists a deterministic ${\bf W} \in
  \mathcal{O}(d)$ such that, with probability at least $1 - 3\eta$,
\begin{equation}\label{conc_x-xhat}
\Bigl| \|\hat{\bf X}-{\bf XW}\|_F-C({\bf X}) \Bigr| \leq \frac{C d^{3/2}
\log{(n/\eta)}}{C(\mathbf{X}) \sqrt{ \gamma_{1}^{7}(\mathbf{P}) \delta(\mathbf{P})}}
\end{equation}
where $C(\mathbf{X})$ is a function of $\mathbf{X}$ given by
\begin{equation*}
C(\mathbf{X}) = \sqrt{\mathrm{tr} \,\, 
    \mathbf{S}_{\mathbf{P}}^{-1/2} \mathbf{U}_{\mathbf{P}}^{T} \mathbb{E}[(\mathbf{A} -
    \mathbf{P})^{2}] \mathbf{U}_{\mathbf{P}}
    \mathbf{S}_{\mathbf{P}}^{-1/2}} 
\end{equation*}
and is bounded from above by $\sqrt{d\gamma_2^{-1}(\mathbf{P})}$. 
Furthermore, under the conditions in
Assumption~\ref{eigengap_assump}, $C(\mathbf{X})$ remains bounded away
from zero as $n \rightarrow \infty$.
\end{theorem}

% \begin{remark} We observe that $C(\mathbf{X})$ is a function of the
%   unknown latent position $\mathbf{X}$. Nevertheless, $C(\mathbf{X})$ can
%   be consistently estimated, as $n \rightarrow \infty$, by the
%   following function $C(\hat{\mathbf{X}})$ of the estimated latent
%   positions $\hat{\mathbf{X}}$
%   \begin{equation*}
%     C(\hat{\mathbf{X}}) = \sqrt{\mathrm{tr} \,\, 
%       \mathbf{S}_{\mathbf{A}}^{-1/2} \mathbf{U}_{\mathbf{A}}^{\top}
%       \hat{\mathbf{D}} \mathbf{U}_{\mathbf{A}} \mathbf{S}_{\mathbf{A}}^{-1/2}}
%   \end{equation*}
%   where $\hat{\mathbf{D}}$ is the diagonal matrix whose diagonal
%   entries are given by
%   \begin{equation*}
%     \hat{\mathbf{D}}_{ii} = \sum_{j \not = i} \langle \hat{X}_i,
%     \hat{X}_j \rangle (1 - \langle \hat{X}_i, \hat{X}_j \rangle).
%   \end{equation*}
% \end{remark}

In the above theorem, 
$\mathbf{U}_{\mathbf{P}} \mathbf{S}_{\mathbf{P}}
\mathbf{U}_{\mathbf{P}}^{\top} = \mathbf{P}$ is the eigendecomposition of $\mathbf{P}$ with
$\mathbf{S}_{\mathbf{P}}$ the $d \times d$ matrix of non-zero
eigenvalues of $\mathbf{P}$. As a corollary of Theorem~\ref{thm:conc_Xhat_X}, we obtain the
following. 
\begin{corollary}
  \label{cor:1}
  Let $\{\mathbf{X}_n\}$ be a sequence of latent positions and suppose
  that the sequence of matrices $\{\mathbf{P}_n\}$ where $\mathbf{P}_n
  = \mathbf{X}_n \mathbf{X}_n^{T}$ satisfies the condition of
  Assumption~\ref{eigengap_assump}. Then there exists a deterministic
  sequence of orthogonal matrices $\mathbf{W}_n$ such that 
  \begin{equation*}
    \|\hX_n - \mathbf{X}_n \mathbf{W}_n \|_F - C(\mathbf{X}_n) 
    \overset{\mathrm{a.s.}}{\longrightarrow} 0
  \end{equation*}
  Furthermore, suppose that the rows of
  $\mathbf{X}_n = [X_1 \mid X_2 \mid \cdots \mid X_n]^{\top}$ are
  sampled according to a distribution $F$ for which the second order
  moment matrix $\mathbb{E}[X_i X_i^{\top}]$ is of rank $d$ with $d$
  distinct eigenvalues. Let $\mu_{F} =
\mathbb{E}[X_1]$ and $\Delta_{F} = \mathbb{E}[X_1 X_1^{\top}]$. Then
  \begin{equation*}
  \|\hX_n - \mathbf{X}_n \mathbf{W}_n \|_F - \sqrt{\mathrm{tr} \Delta_{F}^{-1} \Bigl(\mathbb{E}[X_1 X_1^{\top}
  (X_1^{\top} 
\mu_{F} - X_1^{\top} \Delta_{F} X_1)] \Bigr)
\Delta_{F}^{-1}} \, 
\overset{\mathrm{a.s.}}{\longrightarrow} 0. 
\end{equation*}
\end{corollary}
\begin{remark}
  When the rows of $\mathbf{X}_n$ are
  sampled according to a distribution $F$ satisfying the distinct
  eigenvalues assumption, then by the strong law of large numbers, the
  $\{\mathbf{X}_n\}$ satisfies the condition of
  Assumption~\ref{eigengap_assump} for all but a finite number of
  indices $n$. We then have
\begin{equation*}
C(\mathbf{X}_{n}) = \sqrt{\mathrm{tr} \,\, 
    \mathbf{S}_{\mathbf{P}_n}^{-1/2} \mathbf{U}_{\mathbf{P}_n}^{T} \mathbb{E}[(\mathbf{A}_n -
    \mathbf{P}_n)^{2}] \mathbf{U}_{\mathbf{P}_n}
    \mathbf{S}_{\mathbf{P}_n}^{-1/2}} = \sqrt{\mathrm{tr} \mathbf{X}_n^{T} \mathbf{D}_n \mathbf{X}_n (\mathbf{W}_n \mathbf{S}_{\mathbf{P}_n}^{-1} \mathbf{W}_n^{T})^2}
\end{equation*}
where $\mathbf{W}_n$ is the orthogonal matrix such that
$\mathbf{U}_{\mathbf{P}_n} \mathbf{S}_{\mathbf{P}_n}^{1/2}
\mathbf{W}_n^{T} = \mathbf{X}_n$
and $\mathbf{D}_n$ is the diagonal matrix whose diagonal elements are
$\mathbf{D}_{ii} = \sum_{j \not = i} \langle X_i, X_j \rangle (1 -
\langle X_i, X_j \rangle)$.
By the law of large numbers,
$n^{-1} \mathbf{W}_n \mathbf{S}_{\mathbf{P}_n}^{-1} \mathbf{W}_n^{T} =
n^{-1} (\mathbf{X}^{T} \mathbf{X}_n)^{-1}$
converges to $(\mathbb{E}[X_1 X_1^{\top}])^{-1}$ almost
surely. Furthermore,
$$n^{-2} \mathbf{X}^{T}_n \mathbf{D}_n \mathbf{X}_n = n^{-2}
\sum_{i=1}^{n} \sum_{j \not = i} X_i X_i^{\top} (\langle X_i, X_j
\rangle - X_i X_j X_j^{\top} X_i^{\top})$$
which converges to $\mathbb{E}[X_1 X_1^{\top}
  (X_1^{\top} 
\mu_{F} - X_1^{\top} \Delta_{F} X_1)]$
almost surely. Corollary~\ref{cor:1} provides
the first known distributional result for $\|\hat{\mathbf{X}}_n
- \mathbf{X}_n
\mathbf{W}_n\|_{F}$ in the setting where the rows of
$\mathbf{X}_n$
are independent and identically distributed with distribution $F$.
In this context the corollary complements the result of
\cite{athreya2013limit} wherein it is shown that individual residuals
$\hat{X}_i
-
X_i$ converge to a mixture of multivariate normals; more precisely,
for any fixed $i$,
\begin{equation*}
    \mathbb{P}\Bigl\{\sqrt{n}( \mathbf{W}_n \hat{X}_i - X_i ) \leq z\Bigr\}
\rightarrow  \int 
\Phi(z, \Delta_{F}^{-1}\mathrm{E}[X_j X_j^\top (X_j^{\top} x - X_j^{\top} x x^{\top} X_j)]
 \Delta_{F}^{-1}) \,\, \mathrm{d} F(x)
\end{equation*}
where $\Phi(\cdot, \Sigma)$ denotes the cumulative distribution
function for a multivariate normal with mean $0$ and covariance matrix
$\Sigma$.
\end{remark}

\section{Main results}
\label{sec:main-results}
We present in this section test procedures for testing the hypothesis
of equality (up to rotation) and equality up to scaling. The test
procedure for the hypothesis of equality up to diagonal transformation
is postponed to Section~\ref{sec:extensions} as its theoretical
properties depends on additional assumptions regarding the underlying latent positions
that are unnecessary for our current purpose. 
\subsection{Equality case}
\label{sec:equality}
The first result is
concerned with finite sample and asymptotic properties of a test for
the null hypothesis $H_0: {\bf X}_n \upVdash {\bf Y}_n$ against the
alternative $H_a: {\bf X}_n \nupVdash {\bf Y}_n$, for both the finite
sample case of a fixed pair of latent positions ${\bf X}_n$ and
${\bf Y}_n$ and the asymptotic case of a sequence of latent positions
$\{{\bf X}_n, {\bf Y}_n\}$, $n \in \mathbb{N}$. Before stating the
result, however, we need to present a definition that adapts the
classical notion of consistency to our semiparametric graph inference
setting. Indeed, for the graph testing problems we address, the
parameter dimension grows as the sample size grows and thus motivate
our consideration for consistency of a sequence of hypothesis tests.
We state this definition for the case of testing whether the latent
positions are equal (up to rotation); its adaptation for the scaling
and diagonal tests is clear.

\begin{definition}\label{consistency}
  Let $\mathbf{X}_n$, $\mathbf{Y}_n$ in $\mathbb{R}^{n \times d}$, $n
  \in \mathbb{N}$, be given. A test statistic $T_n$ and associated
  rejection region $R_n$ to test the null hypothesis
  \begin{align*} 
    H^n_0: \,  {\bf X}_n \upVdash {\bf Y}_n \quad
    \textrm{ against } \quad H^n_a: \, {\bf X}_n \nupVdash {\bf Y}_n 
  \end{align*}
  is a {\em consistent, asymptotically level $\alpha$ test} 
  if for any $\eta>0$, there exists $n_0 = n_0(\eta)$ such that 
\begin{enumerate}[(i)]
\item If $n>n_0$ and $H_a^n$ is true, then $P(T_n \in R_n)>1-\eta$
\item If $n > n_0$ and $H_0^n$ is true, then $P(T_n \in R_n) \leq \alpha + \eta$
\end{enumerate}
\end{definition}

% We now present our test procedure. Our test statistic $T_n$ is a scaled version of
% \begin{equation*}
%  \min\limits_{{\bf W} \in \mathcal{O}(d)}
% \|\hat{{\bf X}}_n{\bf W}-\hat{{\bf Y}}_n\|_F. 
% \end{equation*}
% If it is sufficiently small, we do not reject; and if it is larger
% than a constant for which we provide an upper bound, then we
% reject. 

We then have the following result. 
\begin{theorem}
  \label{thm:identity}
  For each fixed $n$, consider the
  hypothesis test
  \begin{equation*}
    H^{n}_0: {\bf X}_n \upVdash {\bf Y}_n \quad \textrm{ versus } \quad
    H^{n}_a: {\bf X}_n \nupVdash {\bf Y}_n
  \end{equation*}
  where ${\bf X}_n$ and ${\bf Y}_n$ $\in \mathbb{R}^{n \times d}$ are
  matrices of latent positions for two random dot product graphs. Let
  $\hat{{\bf X}}_n$ and $\hat{{\bf Y}}_n$ be the adjacency spectral
  embeddings of ${\bf A}_n\sim \mathrm{Bernoulli}({\bf X}_n{\bf
    X}_n^T)$ and ${\bf B}_n \sim \mathrm{Bernoulli}({\bf Y}_n{\bf
    Y}_n^T)$, respectively. Define the test statistic $T_n$ as follows:
  \begin{equation}
    \label{eq:9}
    T_n=\frac{\min\limits_{{\bf W} \in \mathcal{O}(d)} 
      \|\hat{{\bf X}}_n{\bf W}-\hat{{\bf Y}}_n\|_F}
    {\sqrt{d\gamma^{-1}_2(\mathbf{A}_n)}+ \sqrt{d \gamma^{-1}_2(\mathbf{B}_n)}}.
  \end{equation}
  Let $\alpha \in (0,1)$ be given. Then for all $C > 1$, if the
  rejection region is 
$R:=\left\{t \in \mathbb{R}: t\geq C \right\}$, 
then there exists an
$n_1 = n_1(\alpha, C) \in \mathbb{N}$ such that for all $n \geq n_1$, the
test procedure with $T_n$ and rejection region $R$ is an at most level
$\alpha$ test, i.e., for all $n \geq n_1$, if $\mathbf{X}_n
\upVdash \mathbf{Y}_n$, then 
\begin{equation*}
  \mathbb{P}(T_n \in R) \leq \alpha.
\end{equation*}
%if $\{{\bf X}_n, {\bf Y}_n\}$, $n \in
%\mathbb{N}$, is a sequence of pairs of rank $d$ latent positions
%satisfying the conditions in Assumption~\ref{eigengap_assump} %and, in addition, if
Furthermore,
%\begin{equation*}
% \min\limits_{{\bf W} \in \mathcal{O}(d)}\| {\bf X}_n{\bf W}-{\bf Y}_n\| 
 %\rightarrow \infty
%\end{equation*}
consider the sequence of latent positions $\{{\bf X}_n\}$ and 
  $\{{\bf Y}_n\}$, $n \in \mathbb{N}$,
  satisfying Assumption~\ref{eigengap_assump} and denote by $d_n$ the
  quantity 
\begin{equation*}
  d_n := \min\limits_{{\bf W} \in \mathcal{O}(d)} \| {\bf X}_n{\bf W}-{\bf
      Y}_n \|.
  \end{equation*} 
  Suppose $d_n \neq 0$ for infinitely many $n$.  Let $t_1=\min\{k>0:
  d_k>0\}$ and sequentially define $t_n=\min\{k>t_{n-1}: d_k>0\}$.
  Let $b_n=d_{t_n}$.  If $\liminf b_n = \infty$, then this test
  procedure is consistent in the sense of Definition~\ref{consistency}
  over this sequence of latent positions.
\end{theorem}
\begin{remark} This result and its analogues for the scaling and
  diagonal hypotheses do not require that ${\bf A}_n$ and ${\bf B}_n$
  be independent for any fixed $n$, nor that the sequence of pairs
  $({\bf A}_n, {\bf B}_n)$, $n \in \mathbb{N}$, be independent. In
  addition, the requirement that $\liminf b_k=\infty$ can be weakened
  somewhat.  Specifically, consistency is achieved as long
  as $$\liminf_{n \rightarrow \infty} \Bigl(\| \mathbf{X}_n\mathbf{W}
  -\mathbf{Y}_n \|_{F} - C(\mathbf{X}_n) - C(\mathbf{Y}_n)\Bigr) > 0.$$
\end{remark}

\subsection{Scaling case}
\label{sec:scaling}
For the scaling case, let $\mathcal{C}=\mathcal{C}(\mathbf{Y}_n)$
denote the class of all positive constants $c$ for which all the
entries of $c^2 \mathbf{Y}_n \mathbf{Y}_n^T$ belong to the unit
interval. We wish to test the null hypothesis
$H_0 \colon \mathbf{X}_n \upVdash c_n \mathbf{Y}_n$ for some
$c_n\in \mathcal{C}$ against the alternative
$H_a \colon \mathbf{X}_n \nupVdash c_n\mathbf{Y}_n$ for any
$c_n \in \mathcal{C}$. In what follows below, we will only write
$c_n>0$, but will always assume that $c_n \in \mathcal{C}$, since the
problem is ill-posed otherwise. The test statistic $T_n$ is now a
simple modification of the one used in Theorem~\ref{thm:identity}: for
this test, we compute a Procrustes distance between scaled adjacency
spectral embeddings for the two graphs. 
\begin{theorem}
  \label{thm:2}
  For each fixed $n$, consider the
  hypothesis test
  \begin{align*}
    H^{n}_0 & \colon {\bf X}_n \upVdash c_n{\bf Y}_n \quad \text{for some $c_n > 0$}
    \textrm{ versus } \\ 
    H^{n}_a & \colon {\bf X}_n \nupVdash c_n{\bf Y}_n \quad \text{for all $c_n > 0$}
  \end{align*}
  where ${\bf X}_n$ and ${\bf Y}_n$ $\in \mathbb{R}^{n \times d}$ are
  latent positions for two random dot product graphs with adjacency
  matrices $\mathbf{A}_n$ and $\mathbf{B}_n$, respectively. % Let
  % $\hat{{\bf X}}_n$ and $\hat{{\bf Y}}_n$ be the adjacency spectral
  % embeddings of ${\bf A}_n\sim \mathrm{Bernoulli}({\bf X}_n{\bf
  %   X}_n^T)$ and ${\bf B}_n \sim \mathrm{Bernoulli}({\bf Y}_n{\bf
  %   Y}_n^T)$, respectively. 
  Define the test statistic $T_n$ as follows:
  \begin{equation}
    \label{eq:8}
    T_n=\frac{\min\limits_{{\bf W} \in \mathcal{O}(d)} 
      \|\hat{{\bf X}}_n{\bf W}/\|\hX_n\|_{F} - \hat{{\bf
          Y}}_n/\|\hat{\mathbf{Y}}_n\|_{F} \|_{F}}
    {2 \sqrt{d \gamma^{-1}_2(\mathbf{A}_n)}/\|\hX_n\|_{F}+ 2\sqrt{d
        \gamma^{-1}_2(\mathbf{B}_n)}/\|\hat{\mathbf{Y}}_n \|_{F}}.
  \end{equation}
  Let $\alpha \in (0,1)$ be given. Then for all $C
  > 1$, if the rejection region is $R:=\left\{t \in \mathbb{R}: t\geq C \right\}$, 
  then there exists an $n_1 = n_1(\alpha, C) \in \mathbb{N}$ such that
  for all $n \geq n_1$, the test procedure with $T_n$ and rejection
  region $R$ is an at most level $\alpha$ test.
%, i.e., for all $n \geq
%  n_1$, if $\mathbf{X}_n \upVdash c \mathbf{Y}_n$ for some $c > 0$,
%  then
%\begin{equation*}
%  \mathbb{P}(T_n \in R) \leq \alpha.
%\end{equation*}
  Furthermore, consider the sequence of latent position $\{{\bf X}_n\}$ and 
  $\{{\bf Y}_n\}$, $n \in \mathbb{N}$,
  satisfying Assumption~\ref{eigengap_assump} and denote by $d_n$ the quantity
  \begin{equation}\label{eq:scaling_alternative}
  d_n := \frac{ \min\limits_{{\bf W} \in \mathcal{O}(d)}\| {\bf X}_n{\bf W}/\|{\bf
      X}_n\|_{F} - {\bf Y}_n/\|{\bf Y}_n\|_{F}
    \|_{F}}{1/\|\mathbf{X}_n\|_{F} +
    1/\|\mathbf{Y}_n\|_{F}} =   \frac{ \min\limits_{{\bf W} \in \mathcal{O}(d)}
    \| {\bf X}_n \|\mathbf{Y}_n \|_{F} {\bf W}
      - {\bf Y}_n \|\mathbf{X}_n \|_{F}
    \|_{F}}{\|\mathbf{X}_n\|_{F} +
    \|\mathbf{Y}_n\|_{F}}
\end{equation}
Suppose $d_n \neq 0$ for infinitely many $n$.  Let $t_1=\min\{k>0:
d_k>0\}$ and sequentially define $t_n=\min\{k>t_{n-1}: d_k>0\}$.  Let
$b_n=d_{t_n}$.  If $\liminf b_n = \infty$, then this test procedure is
consistent in the sense of Definition~\ref{consistency} over this
sequence of latent positions.
%in Definition \eqref{consistency},
% namely, for any
%$\beta > 0$ there exists a $n_2(\alpha, \beta, C)$ such that $\inf_{n
%  \geq n_2} \mathbb{P}(T_n \in R) \geq 1 - \beta$.
\end{theorem}
\begin{remark}
  We remark that the collection of alternatives in
  Eq. \eqref{eq:scaling_alternative} is effectively those latent
  positions $\mathbf{X}_n$ and $\mathbf{Y}_n$ which, after normalization by their
  Frobenius norms, remain far enough apart as $n \rightarrow \infty$. Indeed,
  the denominator of our test statistic converges to
  zero, so we require that the numerator does not become small too
  quickly. The terms $d \gamma^{-1}_2(\mathbf{A}_n)$ and $d
    \gamma^{-1}_2(\mathbf{B}_n)$ are bounded from above, in the limit, 
  by fixed constants and we can replace them by 
  $1$ to obtain an equivalent class of
  alternatives.
\end{remark}

\section{Experiments}
\subsection{Simulations}
\label{sec:simulations}
\begin{algorithm}[htbp]
\caption{Bootstrapping procedure for the test $\mathbb{H}_0 \colon \mathbf{X} \upVdash \mathbf{Y}$.} 
\label{bootstrap-simple} 
\begin{algorithmic}[1]
\Procedure{Bootstrap}{$\mathbf{X}, T, bs$} \Comment{Returns the p-value associated with $T$.}
\State $d \gets \mathrm{ncol}(\mathbf{X})$ \Comment{Set $d$ to be the number of columns of $\mathbf{X}$.}
\State $\mathcal{S}_{X} \gets \emptyset$
\For{$b \gets 1 \colon bs$}
   \State $\mathbf{A}_b \gets \mathrm{RDPG}(\hat{\mathbf{X}}); \quad \mathbf{B}_b \gets \mathrm{RDPG}(\hat{\mathbf{X}})$
   \State $\hat{\mathbf{X}}_b \gets \mathrm{ASE}(\mathbf{A}_b, d); \quad \hat{\mathbf{Y}}_b \gets \mathrm{ASE}(\mathbf{B}_b, d)$
   \State $T_b \gets \min_{\mathbf{W}} \| \hat{\mathbf{X}}_{b} - \hat{\mathbf{Y}}_{b} \mathbf{W} \|_{F}; \qquad \mathcal{S}_{X} \gets \mathcal{S}_X \cup T_b$    
\EndFor
\State \Return $p \gets (|\{s \in \mathcal{S}_X \colon s \geq T\}| + 0.5)/bs$ \Comment{Continuity correction.} 
\EndProcedure \\
\State $\hat{\mathbf{X}} \gets \mathrm{ASE}(\mathbf{A}, d)$ \Comment{The embedding dimension $d$ is assumed given.}
\State $\hat{\mathbf{Y}} \gets \mathrm{ASE}(\mathbf{B}, d)$
\State $T \gets \min_{\mathbf{W}} \|\hat{\mathbf{X}} - \hat{\mathbf{Y}} \mathbf{W} \|_{F}$
\State $p_{X} \gets \mathrm{Bootstrap}(\hat{\mathbf{X}}, T, bs)$ \Comment{The number of bootstrap samples $bs$ is assumed given.}
\State $p_{Y} \gets \mathrm{Bootstrap}(\hat{\mathbf{Y}}, T, bs)$ 
\State $p = \max\{p_X, p_Y\}$ \Comment{Returns the maximum of the two p-values.}
\end{algorithmic}
\end{algorithm}

In this section, we illustrate the test
procedure of Section~\ref{sec:main-results} through several simulated data
examples. We first consider the problem of testing the null
hypothesis $H_0 \colon \mathbf{X}_n \upVdash \mathbf{Y}_n$ against the
alternative hypothesis $H_{A} \colon \mathbf{X}_{n} \nupVdash
\mathbf{Y}_n$. We consider random graphs generated according to two
stochastic blockmodels with the same block
membership probability vector $\bm{\pi}$ but different block
probability matrices. Define $\mathbf{B}_{\epsilon}$ for $\epsilon
\geq 0$ by
\begin{equation}
  \label{eq:16}
  \mathbf{B}_{\epsilon} = \begin{bmatrix} 0.5 + \epsilon & 0.2 \\ 0.2
    & 0.5 + \epsilon \end{bmatrix}.
\end{equation}
We then test, for a given $\epsilon > 0$, the hypothesis $H_0 \colon
\mathbf{X}_n \upVdash \mathbf{Y}_{n}^{(\epsilon)}$ against 
$H_{A} \colon \mathbf{X}_n \nupVdash 
\mathbf{Y}_{n}^{(\epsilon)}$ where $\mathbf{X}_n$ corresponds to
$\mathbf{B}_0$ and $\mathbf{Y}_n^{(\epsilon)}$ corresponds to
$\mathbf{B}_{\epsilon}$.  We evaluate the performance of the test
procedure by estimating the level and power of the test statistic for
various choices of $n \in \{100, 200, 500, 1000 \}$ and $\epsilon \in
\{0, 0.05, 0.1, 0.2\}$ through Monte Carlo simulation. The
significance level is set to $\alpha = 0.05$ and the rejection
regions are specified via one of two approaches, namely (1) a
bootstrap procedure based on the 
the estimated latent positions $\hat{\mathbf{X}}_n$ and
$\hat{\mathbf{Y}}_n$ (see Algorithm~\ref{bootstrap-simple}) and (2) $\{T > 1\}$ as dictated by the
asymptotic theory. The results are given in
Table~\ref{tab:bootstrap_identity}. To keep the vertex set fixed and 
aligned, the block membership vector is sampled once in 
each Monte Carlo replicate. Table~\ref{tab:bootstrap_identity} indicates that the test has good
power and is indeed asymptotically level $\alpha$. The rejection
regions computed using bootstrap resampling are generally less
conservative than those specified via the asymptotic
theory. Nevertheless, the theoretical rejection regions exhibit power
even for moderate values of $n$ such as $n = 200$.

\begin{table}[!t]
  \footnotesize
  \centering
\begin{tabular}{ccccccccc}
 & \multicolumn{2}{c}{$\epsilon = 0$} 
 & \multicolumn{2}{c}{$\epsilon = 0.05$} &
 \multicolumn{2}{c}{$\epsilon = 0.1$} & \multicolumn{2}{c}{$\epsilon
   = 0.2$} \\ 
 $n$ & bootstrap & theoretical & bootstrap & theoretical & bootstrap
 & theoretical & bootstrap & theoretical \\ \midrule 
 $100$ & $0.08$ & $0$ & $0.11$ & $0$  & $0.31$ & $0$ & $0.99$ & $0.13$ \\
 $200$ & $0.07$ & $0$ & $0.22$ & $0$  & $0.96$ & $0$ & $1$ & $0.98$ \\
 $500$ & $0.06$ & $0$ & $0.97$ & $0$  & $1$ & $0$ & $1$ & $1$ \\
 $1000$& $0.05$ & $0$ & $1$ & 0 & $1$ & $1$ & $1$ & $1$ 
% \midrule
\end{tabular}
\caption{Power estimates for testing the null hypothesis $\mathbf{X}_n
  \upVdash \mathbf{Y}_n$ at a significance level of $\alpha =
  0.05$. The rejection regions are specified via two methods (1) the
  asymptotic theoretical rejection region and (2) bootstrap
  permutation with $B = 200$ bootstrap samples. Each estimate of power is 
  based on $1000$ Monte Carlo replicates.}
\label{tab:bootstrap_identity}
\end{table}

 \begin{table}[!t]
  \footnotesize
  \centering
\begin{tabular}{ccccccccc}
 & \multicolumn{2}{c}{$\epsilon = 0$} 
 & \multicolumn{2}{c}{$\epsilon = 0.1$} &
 \multicolumn{2}{c}{$\epsilon = 0.2$} & \multicolumn{2}{c}{$\epsilon
   = 0.4$} \\ 
 $n$ & bootstrap & theoretical & bootstrap & theoretical & bootstrap
 & theoretical & bootstrap & theoretical \\ \midrule 
 $100$ & $0.08$ & $0$ & $0.08$ & $0$  & $0.19$ & $0$ & $0.87$ & $0$ \\
 $200$ & $0.06$ & $0$ & $0.15$ & $0$  & $0.61$ & $0$ & $1$ & $0$ \\
 $500$ & $0.05$ & $0$ & $0.62$ & $0$  & $1$ & $0$ & $1$ & $1$ \\
 $1000$& $0.04$ & $0$ & $1$ & $0$ & $1$ & $0$ & $1$ & $1$ 
% \midrule
\end{tabular}
\caption{Power estimates for testing the null hypothesis $\mathbf{X}_n
  \upVdash c_n \mathbf{Y}_n$ for some $c_n > 0$ at a significance level of $\alpha =
  0.05$. The rejection regions are specified via two methods (1) the
  asymptotic theoretical rejection region and (2) bootstrap
  permutation with $B = 200$ bootstrap samples. Each estimate of power is 
  based on $1000$ Monte Carlo replicates.}
\label{tab:bootstrap_scale}
\end{table}
 
We next consider the hypothesis test $H_0 \colon \mathbf{X}_n \upVdash c_n
\mathbf{Y}_n$ for some $c_n > 0$ against the alternative
$H_{A} \colon \mathbf{X}_n \nupVdash c_n \mathbf{Y}_n$ for any $c_n
> 0$. We again employ the model
specified in Eq.~\eqref{eq:16}. 
% \begin{figure}[htbp]
% %    \captionsetup[subfig]{labelformat=empty}
%     \centering
%     \includegraphics[width=0.8\textwidth]{testc.pdf}
% %    \subfloat{
% %    \includegraphics[width=0.8\textwidth]{test1_n1000.pdf}}
%   \caption{Density estimate for the test statistic when testing $H_0
%     \colon \mathbf{X}_n \upVdash c_n \mathbf{Y}_n$ for some $c_n > 0$
%     against the alternative $H_{A} \colon \mathbf{X}_n \nupVdash c_n
%     \mathbf{Y}_n$ for all $c_n > 0$.}
%   \label{fig:testc}
%   \end{figure}
The results are presented in Table~\ref{tab:bootstrap_scale}. Once
again, the significance level is set to $\alpha = 0.05$ and the
rejection regions are specified via one of two approaches, namely (1)
bootstrap resampling from the estimated latent positions
$\hat{\mathbf{X}}_n$ and $\hat{\mathbf{Y}}_n$ similar to
Algorithm~\ref{bootstrap-simple}) and (2) $\{T > 1\}$ as dictated by
the asymptotic theory. We observe that the power of the test is
estimated to be roughly $0.19$ for $n = 100$ and $\epsilon = 0.2$,
which is significantly smaller than the corresponding estimate of
$0.99$ in Table~\ref{tab:bootstrap_identity}, even though the random
graphs models are identical. This is consistent with the notion that
the null hypothesis considered in Table~\ref{tab:bootstrap_identity}
is a single element of the hypothesis space in
Table~\ref{tab:bootstrap_scale}. For this setup, the theoretical
rejection region as specified in Theorem~\ref{thm:2} exhibits power
for moderate values of $n = 500$ and $\epsilon = 0.4$.
  
 \begin{figure}[tb!]
   \centering
   \includegraphics[width=0.5\textwidth]{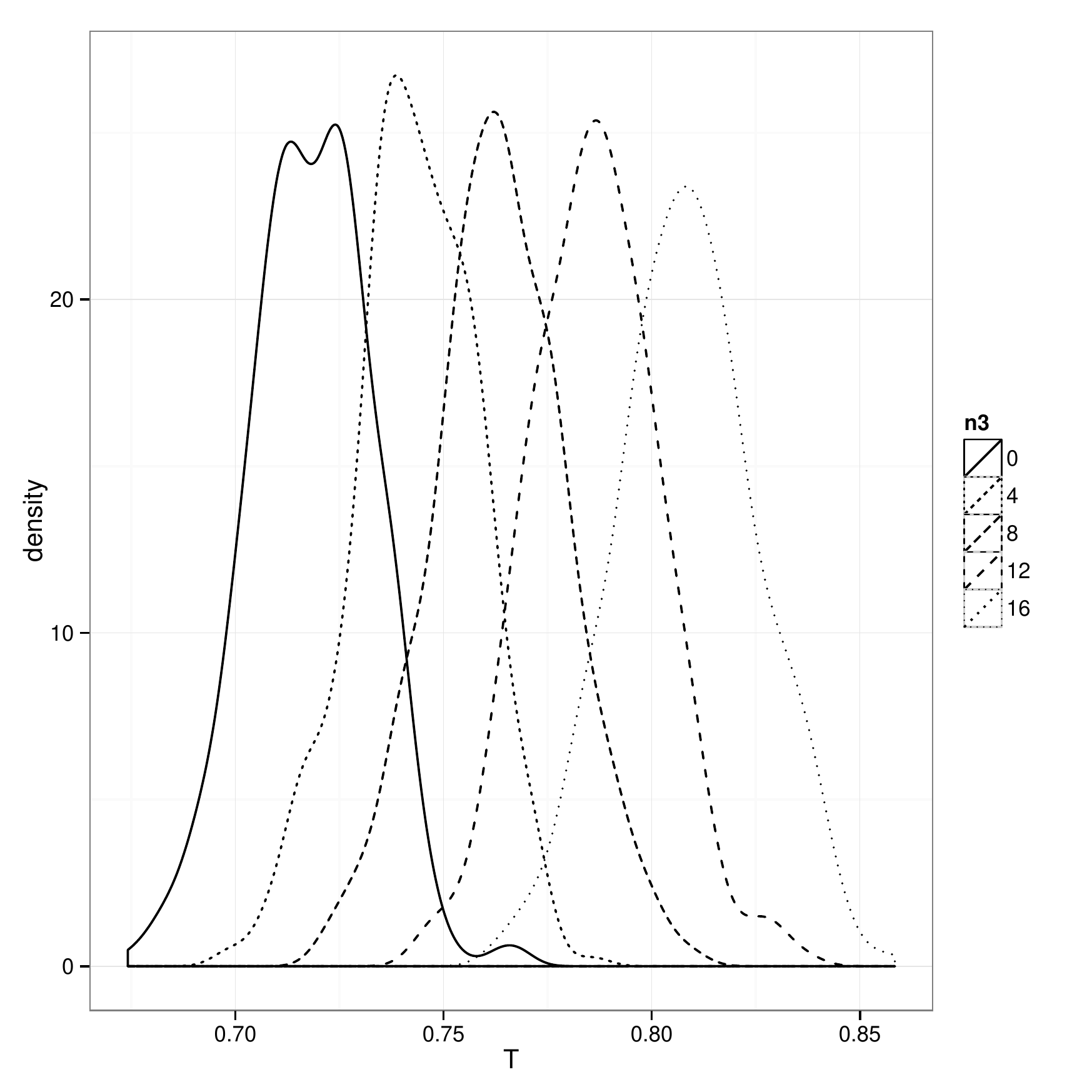}
   % % \subfloat{
   % % \includegraphics[width=0.8\textwidth]{test1_n1000.pdf}}
   \caption{Density estimate (based on $500$ Monte Carlo replicates) for the test statistic to detect
     the emergence of a community of size $n_3 \in \{0,4,\dots,16\}$
     in a graph on $n = 800$ vertices. Bootstrap estimates of the
     critical values for $\alpha = 0.05$ yield power estimates of
     $0.57$ for $n_3 = 4$, $0.92$ for $n_3 = 8$, and $1.0$ for
     $n_3 = 12$ and $n_3 = 16$.}
   % against $H_{A} \colon \mathbf{X}_n \nupVdash
   % \mathbf{Y}_n$. The theoretical rejection region is
%     $T > 1$.}
   \label{fig:community}
 \end{figure}
 
As the last example, we consider the problem of detecting the emergence
of a new community in a graph. This example illustrates, albeit rather
naively, the applicability of the proposed hypothesis test to anomaly
detection in a time series of graphs. Let
$\mathbf{B}_0$ and $\mathbf{B}_1$ be block probability matrices
defined by
\begin{equation*}
  \mathbf{B}_1 \begin{pmatrix}
0.34 & 0.25  \\
0.25 & 0.25  \\
\end{pmatrix};
  \qquad 
  \mathbf{B}_2 = \begin{pmatrix}
    0.34 & 0.25 & 0.16 \\
    0.25 & 0.25 & 0.25 \\
    0.16 & 0.25 & 0.34
\end{pmatrix}
\end{equation*}
Graphs generated with block probability matrix $\mathbf{B}_1$ have
two blocks of size $400$ each while graphs with block probability
matrix $\mathbf{B}_2$ have three blocks of size $400 - n_3/2$, $400 -
n_3/2$ and $2n_3$. The results are presented in
Figure~\ref{fig:community} for various values of $n_3 \in
\{0,4,\dots,16\}$.  
 
 \subsection{C. elegans wiring diagram}
 \label{sec:em-c.-elegans}
 We now apply our test procedure to the two neuronal networks of the
 {\em C. elegans} roundworm.  As we remarked earlier in
 \S~\ref{sec:introduction}, the {\em C. elegans} connectome has two
 distinct connection types, chemical synapses and electrical gap
 junctions, and these two synaptic types give rise to two distinct
 brain graphs.  In each connectome, there are 302 total neurons, with
 20 neurons belonging to the phyrangeal nervous system and the
 remaining 282 belonging to the somatic nervous system.  These two
 nervous systems are disjoint in both connectomes, and we focus our
 attention on the larger somatic nervous system.  Moreover, in the
 somatic nervous system there are three neurons that have no synaptic
 connection to other neurons.  After removing these, we are left with
 two graphs: $\mathbf{A}_{c}$ for the chemical synapses and
 $\mathbf{A}_{g}$ for the gap junctions.  Both graphs are on $279$
 vertices with $\mathbf{A}_{c}$ having $6393$ undirected edges and
 graph $\mathbf{A}_{c}$ having $1031$ undirected edges. See
 \cite{varshney11:_struc} for more detailed description of the
 construction of these connectomes.  

 In each connectome, the neurons are classified into three classes
 that correspond roughly to the sensory neurons, interneurons and
 motor neurons, and Table~\ref{table:celegans_data} (reproduced from
 \citet{varshney11:_struc}) summarizes the number of connections
 between the different types of neurons for the chemical and
 electrical wiring graphs. We frame the question of whether these two
 graphs are ``similar" as a two-sample testing problem.  Because the
 two graphs have a significant difference in the number of edges, the
 appropriate null hypothesis is that the generating latent positions
 are equal up to some scaling factor $c$.

\begin{table}[!t]
  \footnotesize
  \centering
  \subfloat[][Gap junction]{
    \begin{tabular}{|c|c|c|c|}
      \hline
      & sensory & inter & motor \\ \hline
      sensory & 108 (42.7\%) & 119 (47.0\%) & 26 (10.3 \%) \\ 
      inter & 119 (14.4\%) & 368 (44.4\%) & 342 (41.3\%) \\
      motor & 26 (3.8\%) & 342 (49.4\%) & 324 (46.8\%) \\ \hline
    \end{tabular}
  } \hfill
  \subfloat[][Chemical synapses]{
    \begin{tabular}{|c|c|c|c|}
      \hline
      & sensory & inter & motor \\ \hline
      sensory & 474 (21.0\%) & 1434 (63.4\%) & 353 (15.6\%) \\ 
      inter & 208 (8.3\%) & 1359 (54.5 \%) & 929 (37.2 \%) \\
      motor  & 30 (1.8 \%) & 275 (16.8 \%) & 1332 (81.4 \%) \\ \hline
    \end{tabular}
  }
  \caption{Numbers of connections between types of neurons in the
    electrical and chemical wiring of {C.elegans}, from \citet{varshney11:_struc}.}
  \label{table:celegans_data}
\end{table}

To carry out the test, we embed each graph as a collection of points
in $\mathbb{R}^{d}$ with $d = 6$. The choice of $d = 6$ is selected
using the automatic dimension selection procedure of
\cite{zhu06:_autom}. Denoting by $\hat{\mathbf{X}}_{c}$ and
$\hat{\mathbf{X}}_{g}$ the resulting embeddings, we compute the test
statistic
\begin{equation*}
  T(\hat{\mathbf{X}}_{c}, \hat{\mathbf{X}}_{g}) = 
  \frac{\min\limits_{{\bf W} \in \mathcal{O}(d)} 
    \|\hat{{\bf X}}_g{\bf W}/\|\hX_g\|_{F} - \hat{{\bf X}}_c/\|\hat{\mathbf{X}}_c\|_{F} \|_{F}}
  {2 C(\hat{\mathbf{X}}_{g})/\|\hX_g\|_{F}+ 
    2 C(\hat{\mathbf{X}}_c)/\|\hat{\mathbf{Y}}_c \|_{F}}
= 1.465,
\end{equation*}
as described in Section~\ref{sec:main-results}.
% We obtain $T(\hat{\mathbf{X}}_{c}, \hat{\mathbf{X}}_{g}) = 1.465$.
To approximate the $p$-value, we modify the bootstrapping procedure in
Algorithm~\ref{bootstrap-simple} and set $T_b$ to the statistic in the
above display. The number of bootstrap samples is set to $bs = 1000$.
The approximate $p$-value associated with the value $T = 1.465$ of the
test statistic is smaller than $ 0.001$.  Hence, we reject the null
and conclude that the two connectomes are sufficiently different, even
up to a density-correcting scaling factor.  The analysis of
\citet{varshney11:_struc}, and in particular the connection
probabilities they provide, as reproduced in
Table~\ref{table:celegans_data} above, appears to support this
conclusion; however, the biological implications of this warrant
further investigation.  We note that there is no general consensus
within the biological community as to how ``similar" the two graphs
are.

\subsection{Neuroimaging data}
\begin{figure}[htbp]
  \centering
  \includegraphics[width=\textwidth]{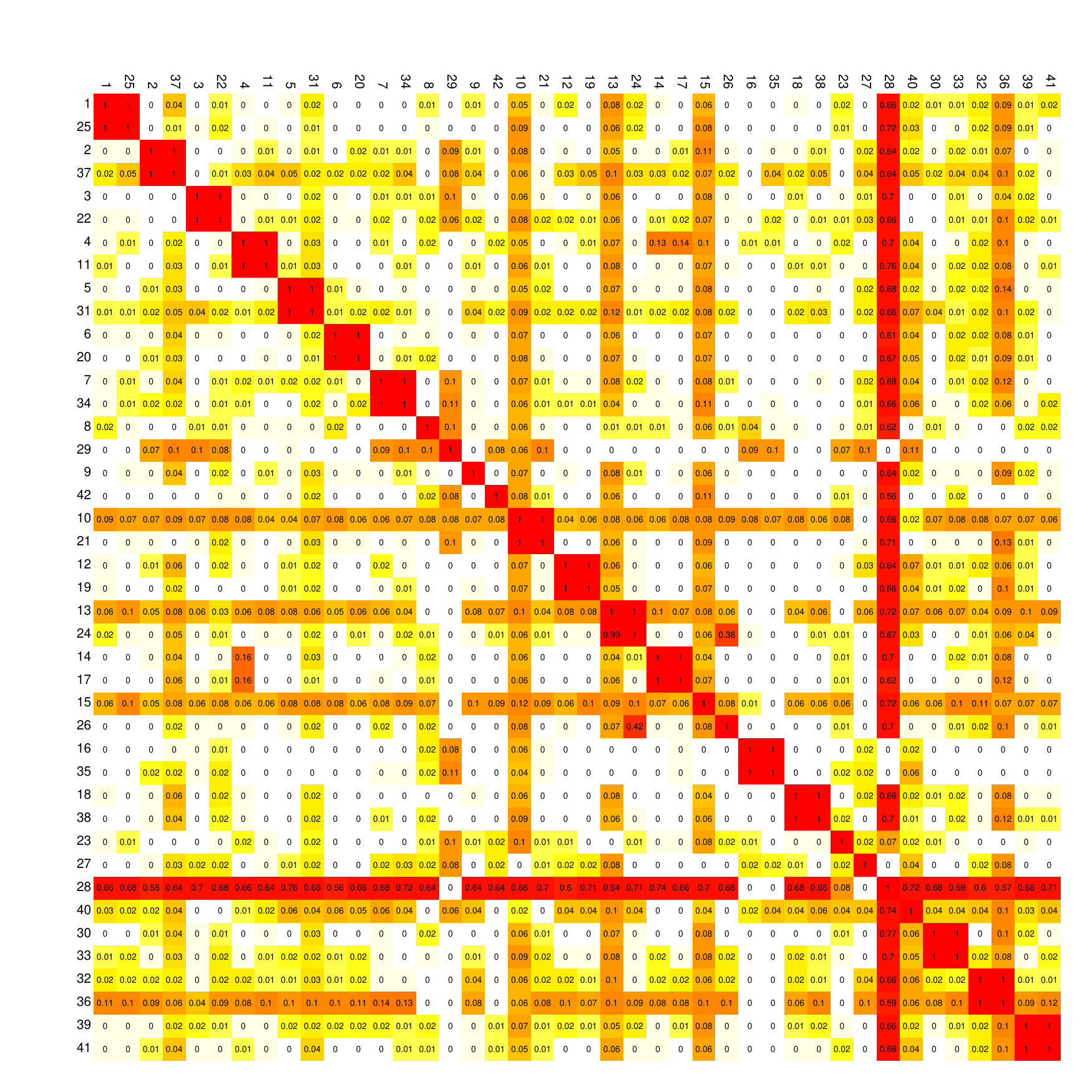}
  \caption{Matrix of p-values (uncorrected) for testing the hypothesis
    $\mathbb{H}_0 \colon \mathbf{X} \upVdash \mathbf{Y}$ for the $42
    \times 41/2$ pairs of graphs generated from the KKI test-retest
    dataset of \citet{landman_KKI}. The labels had been arranged so
    that the pair $(2i-1,2i)$ correspond to scans from the same
    subject. The $p$-values are color coded to vary in intensity from
    white ($p$-value of $0$) to dark red ($p$-value of $1$).}
  \label{fig:kki-small-identity}
\end{figure}

We end this section by applying our test procedure to the test-retest
diffusion MRI data from \citet{landman_KKI}. We recall that, for this
example, the raw data consist of 42 images: namely, one pair of neural
images from each of 21 subjects. These images are generated for the
purpose of evaluating scan-rescan reproducibility of the
magnetization-prepared rapid acquistion gradient echo (MPRAGE) image
protocol. Table 5 from \citet{landman_KKI} indicates that the
variability of MPRAGE is quite small; specifically, the cortical gray
matter, cortical white matter, ventricular cerebrospinal fluid,
thalamus, putamen, caudate, cerebellar gray matter, cerebellar white
matter, and brainstem were identified with mean volume-wise
reproducibility of $3.5\%$, with the largest variability being that of
the ventricular cerebrospinal fluid at $11\%$.  These scans can be
converted into graphs at various scales. We first consider a
collection of small graphs on seventy vertices that are generated from
seventy brain regions and the fibers connecting them.  Given these
graphs, we proceed to investigate the similarities and dissimilarities
between the scans.  We first embed each graph into
$\mathbb{R}^{4}$. We then test the hypothesis of equality up to
rotation with the p-values obtained using the parametric bootstrapping
procedure in Algorithm~\ref{bootstrap-simple}.  The results are
presented in Figure~\ref{fig:kki-small-identity}.
Figure~\ref{fig:kki-small-identity} indicates that, in general, the
test procedure fails to reject the null hypothesis when the two graphs
are for the same subject.  This is consistent with the reproducibility
finding of \citet{landman_KKI}. Furthermore, this outcome is also
intuitively plausible; in addition to failing to reject when two scans
are from the same subject, we also frequently {\em do} reject the null
hypothesis when the two graphs are from scans of different
subjects. Note that our analysis is purely exploratory; as such, we do
not grapple with issues of multiple comparisons here.

Similar results hold when we consider the large graphs generated from
these test-retest data through the MIGRAINE pipeline of
\citet{gray_migraine}.  For each magnetic resonance scan, the MIGRAINE
pipeline generates graphs with roughly $10^7$ vertices and $10^{10}$
edges with the vertices of all the graphs aligned.  Because many of these
voxels are noise (due to the choice of masking employed by the
pipeline), the graphs are then reduced to their largest connected
component. These largest connected components preserve essentially all
white matter voxels and are on the order of $10^5$ vertices
and $10^8$ edges. Bootstrapping the test statistics for these large
graphs present some practical difficulties. Indeed, the
bootstrapping procedure in Algorithm~\ref{bootstrap-simple} requires
generating multiple graphs on the order of $10^5$ vertices. The time
and space complexity for generating a naive matrix representation of
such graphs is $O(n^2)$, where $n$ denotes the number of vertices;
meanwhile, the time and space complexity to generate a sparse
representation of such graphs is $O(m)$ \citep{batagelj:large-graphs}
where $m$ denotes the number of edges. In particular, the space
complexity for each bootstrap sample is prohibitively large for
current commodity computing resources. A more efficient bootstrapping
procedure suitable for large graphs is thus desired.

\begin{algorithm}[t!]
  \caption{Subgraphs bootstrapping procedure for the test $\mathbb{H}_0
    \colon \mathbf{X} \upVdash \mathbf{Y}$.}
  \label{bootstrap-complex}
  \begin{algorithmic}[1]
    \State $\hat{\mathbf{X}} \gets \mathrm{ASE}(\mathbf{A}, d)$
    \Comment{The embedding dimensions $d$ is assumed given.}  \State
    $\hat{\mathbf{Y}} \gets \mathrm{ASE}(\mathbf{B}, d)$ \State $V
    \rightarrow V_1 \cup V_2 \cdots \cup V_R$ \Comment{Partition the
      set of vertices into blocks} \For{$r \gets 1 \colon R$} \State
    $\hat{\mathbf{X}}_r \gets \hat{\mathbf{X}}_{|V_r}$
    \Comment{$\hat{\mathbf{X}}_{r}$ are the rows of $\mathbf{X}$ for
      vertices in $V_r$} \State $\hat{\mathbf{Y}}_{r} \gets
    \hat{\mathbf{Y}}_{|V_r}$
   \State $T_r \gets \min_{\mathbf{W}} \|\hat{\mathbf{X}}_r
    -\hat{\mathbf{Y}}_r \mathbf{W} \|_{F}$ \State $p_{X,r} \gets
    \mathrm{Bootstrap}(\hat{\mathbf{X}}_r, T_r, bs)$ \Comment{Invoke
      the bootstrap procedure in Algorithm~\ref{bootstrap-simple}.}
    \State $p_{Y,r} \gets \mathrm{Bootstrap}(\hat{\mathbf{Y}}_r, T_r,
    bs)$
    \EndFor
    \State $p_{X,\chi} \gets 2 \sum_{r=1}^{R} \log (1/p_{X,r})$ \State
    $p_{Y,\chi} \gets 2 \sum_{r=1}^{R} \log (1/p_{Y,r}) $ \State $p
    \gets \max\{G^{-1}(p_{X,\chi}), G^{-1}(p_{Y,\chi}))\}$
    \Comment{$G$ is the cdf for a $\chi^{2}_{2R}$ random variable.}
  \end{algorithmic}
\end{algorithm}

We propose such a procedure in Algorithm~\ref{bootstrap-complex}. In
Algorithm~\ref{bootstrap-complex}, the vertices of the graphs are
partitioned into $R$ blocks. Suppose for simplicity that each block
contains $n/R$ vertices.  The bootstrapping procedure in
Algorithm~\ref{bootstrap-complex} can then be implemented in time
complexity $O(n^2/R)$ and space complexity $O(n^2/R^2)$. Provided that
$R$ is suitably chosen, this yields a computationally efficient
version of Algorithm~\ref{bootstrap-simple} for large graphs. The
justification behind Algorithm~\ref{bootstrap-complex} is as
follows. Under the null hypothesis of $\mathbf{X} \upVdash
\mathbf{Y}$, any submatrices $\mathbf{X}_r = \mathbf{X}_{|V_r}$ and
$\mathbf{Y}_r = \mathbf{Y}_{|V_r}$ of $\mathbf{X}$ and $\mathbf{Y}$ on
the same collection of rows (indexed by $V_r$) also satisfy
$\mathbf{X}_r \upVdash \mathbf{Y}_r$. Therefore under the null
hypothesis, the induced subgraphs $\mathbf{A}_r \sim
\mathrm{RDPG}(\mathbf{X}_r)$ and $\mathbf{B}_r \sim
\mathrm{RDPG}(\mathbf{Y}_r)$ will yield a value of the test
statistic with a ``large'' p-value. By repeatedly sampling different
induced subgraphs $\mathbf{A}_r$ and $\mathbf{B}_r$ of $\mathbf{A}$
and $\mathbf{B}$, we obtain a
collection of p-values. Assuming that these p-values are independent
(which is the case when no two induced subgraphs overlap), we
can combine them using Fisher's combined probability test
\citep{mosteller48:_quest_answer}. Under the null hypothesis, the
resulting statistic can be approximated by a chi-square distribution
with the appropriate degrees of freedom.

As an illustrative example, we consider the graphs corresponding to
scans $1$, $3$, and $4$; scans $1$ and $3$ coming from the same
subject and scan $4$ from a different subject. The embedding dimension
is chosen to be $50$ while $R$ is chosen so that $n/R \approx
1000$.  For scans $1$ and $3$ from the same subject, the subgraphs
bootstrapping procedure in Algorithm~\ref{bootstrap-complex} yields a
p-value of $0.35$; meanwhile, for scans $1$ and $4$ from different
subjects, the p-value is
$0.00625$. These are consistent with the results for the small graphs
on $70$ vertices and, furthermore, confirm the applicability of our
test procedure to large graphs.

\section{Diagonal transformation case}
\label{sec:extensions}

%\subsection{Scaling Case}
We now consider the case of testing whether the latent positions are
related by diagonal transformation.  
%\subsection{Diagonal Case}
i.e., whether $H_0
\colon \mathbf{X}_n \upVdash \mathbf{D}_n \mathbf{Y}_n$ for some
diagonal matrix $\mathbf{D}_n$. We proceed analogously to the
scaling case in Section~\ref{sec:main-results} by defining the class
$\mathcal{E}=\mathcal{E}(\mathbf{Y}_n)$ to be all positive diagonal
matrices $\mathbf{D}_n \in \mathbb{R}^{n \times n}$ such that
$\mathbf{D}_n \mathbf{Y}_n \mathbf{Y}_n^T \mathbf{D}_n$ has all
entries in the unit interval.  % Once more, we focus on testing the null
% hypothesis $H_0 \colon \mathbf{X}_n \upVdash \mathbf{D}_n
% \mathbf{Y}_n$ for some $\mathbf{D}_n \in \mathcal{E}$ against the
% alternative $H_a \colon \mathbf{X}_n \nupVdash
% \mathbf{D}_n\mathbf{Y}_n$ for any matrix $\mathbf{D}_n \in
% \mathcal{E}$.  
As before, we will
always assume that $\mathbf{D}_n$ belongs to $\mathcal{E}$, even if
this assumption is not explicitly stated.  The test statistic $T_n$ in
this case is again a simple modification of the one used in
Theorem~\ref{thm:identity}. However, for technical reasons, our proof
of consistency requires an additional condition on the minimum
Euclidean norm of each row of the matrices $\mathbf{X}_n$ and
$\mathbf{Y}_n$. To avoid certain technical issues, we impose a
slightly stronger density assumption on our graphs for this test.
These assumptions can be weakened, but at the cost of interpretability.
The assumptions we make on the latent positions, which we summarize
here, are moderate restrictions on the sparsity of the graphs.
\begin{assumption}
  \label{eigengap_assump_diagonal}
  We assume that there exists $d \in \mathbb{N}$ such that for all
  $n$, $\mathbf{P}_n$ is of rank $d$.  Further, we assume that there
  exist constants $\epsilon_1>0$, $\epsilon_2 >0$, $c_0>0$ and
  $n_0(\epsilon_1, \epsilon_2, c) \in \mathbb{N}$ such that for all $n \geq n_0$:
\begin{align}
  \gamma_1(\mathbf{P}_n) &> c_0\\
  \delta(\mathbf{P}_n) &> n^{1/2} (\log{n})^{\epsilon_1} \\ 
  \min_{i} \|X_i\| &>
  \left(\frac{\log{n}}{\sqrt{\delta(\mathbf{P}_n)}}\right)^{1 -
    \epsilon_2}
\end{align}
\end{assumption}
We then have the following result.
\begin{theorem}
  \label{thm:1}
  For each fixed $n$, consider the
  hypothesis test
  \begin{align*}
    H^{n}_0 & \colon {\bf X}_n \upVdash \mathbf{D}_n {\bf Y}_n \quad
    \text{for some diagonal $\mathbf{D}_n \in \mathcal{E}$}
    \textrm{ versus } \\ 
    H^{n}_a & \colon {\bf X}_n \nupVdash \mathbf{D}_n{\bf Y}_n \quad \text{for
      any diagonal $\mathbf{D}_n \in \mathcal{E}$}
  \end{align*}
  where ${\bf X}_n$ and ${\bf Y}_n$ $\in \mathbb{R}^{n \times d}$ are
  matrices of latent positions for two random dot product graphs. 
  For any matrix $\mathbf{Z} \in \mathbb{R}^{n \times d}$, let
  $\mathcal{D}(\mathbf{Z})$ be the diagonal matrix whose diagonal
 entries are the Euclidean norm of the rows of $\mathbf{Z}$ and let
%  \begin{equation*}
%    \mathcal{D}(\mathbf{Z}) = (\mathrm{diag}(\mathbf{Z} \mathbf{Z}^{T}))^{1/2}.
%  \end{equation*}
  $\mathcal{P}(\mathbf{Z})$ be the matrix whose rows are the projection of the rows of
  $\mathbf{Z}$ onto the unit sphere. %  \begin{equation*}
%    \mathcal{P}(\mathbf{Z}) = \mathcal{D}^{-1}(\mathbf{Z}) \mathbf{Z}
%  \end{equation*}
%   For any matrix $\mathbf{M} \in \mathbb{R}^{n \times n}$ with
%   positive entries, define $S(\mathbf{M})$ as follows:
% \begin{equation}\label{eq:defn_S}
% S(\mathbf{M})=\frac{170 d^{3/2}
% \log{(n/\eta)}}{\sqrt{ \gamma_{1}^{7}(\mathbf{M}) \delta(\mathbf{M})}}.
% \end{equation}
We define the test statistic as follows:
  \begin{equation}
    \label{eq:15}
    T_n=\frac{\min\limits_{{\bf W} \in \mathcal{O}(d)} 
      \|\mathcal{P}(\hat{\mathbf{X}}_n) {\bf W} -
     \mathcal{P}(\hat{\mathbf{Y}}_n) \|_{F}}
    {2 \sqrt{d \gamma_{2}^{-1}(\mathbf{A})} \|\mathcal{D}^{-1}(\hX_n)\|_{2}+
 2 \sqrt{d \gamma_{2}^{-1}(\mathbf{B}_n)}\|\mathcal{D}^{-1}(\hat{\mathbf{Y}}_n)\|_{2}}.
  \end{equation}
  where we write
  $\mathcal{D}^{-1}(\mathbf{Z})$ for
  $(\mathcal{D}(\mathbf{Z}))^{-1}$. Note that
  $\|\mathcal{D}^{-1}(\mathbf{Z})\| = 1/(\min_{i} \|Z_i\|)$. 
 
  Let $\alpha \in (0,1)$ be given. Then for all $C
  > 1$, if the rejection region is 
%defined by
  $R:=\left\{t \in \mathbb{R}: t\geq C \right\},$
  then there exists an $n_1 = n_1(\alpha, C) \in \mathbb{N}$ such that
  for all $n \geq n_1$, the test procedure with $T_n$ and rejection
  region $R$ is an at most level-$\alpha$ test.
% , i.e., for all $n \geq n_1$, if $\mathbf{X}_n
%   \upVdash \mathbf{D} \mathbf{Y}_n$ for some diagonal matrix
%   $\mathbf{D}$ with diagonal entries positive and bounded away from $0$, then 
% \begin{equation*}
%   \mathbb{P}(T_n \in R) \leq \alpha.
% \end{equation*}
Furthermore, consider the sequence of latent position $\{{\bf X}_n\}$ and $\{{\bf Y}_n\}$, $n \in \mathbb{N}$,
  satisfying 
  Assumption~\ref{eigengap_assump_diagonal} and denote by $d_n$ the quantity
\begin{equation}
  \label{eq:12}
  d_n := 
\frac{ \min\limits_{{\bf W} \in \mathcal{O}(d)}\| \mathcal{P}({\bf
    X}_n) {\bf W} -
 \mathcal{P}({\bf Y}_n)
 \|_{F}}{\|\mathcal{D}^{-1}(\mathbf{X})\|_{2} +
 \|\mathcal{D}^{-1}(\mathbf{Y})\|_{2}} = D_{\mathcal{P}}(\mathbf{X}_n, \mathbf{Y}_n) 
\end{equation}
% \begin{equation}
%   \label{eq:12}
%   D_{\mathcal{P}}(\mathbf{X}, \mathbf{Y}) \colon = 
% \frac{ \min\limits_{{\bf W} \in \mathcal{O}(d)}\| \mathcal{P}({\bf
%     X}_n) {\bf W} -
%  \mathcal{P}({\bf Y}_n)
%  \|_{F}}{S(\mathbf{P}_n)\|\mathcal{D}^{-1}(\mathbf{X})\|_{F} +
%  S(\mathbf{Q}_n)\|\mathcal{D}^{-1}(\mathbf{Y})\|_{F}} = d_n
% \end{equation}
Suppose $d_n \neq 0$ for infinitely many $n$.  Let $t_1=\min\{k>0:
d_k>0\}$ and sequentially define $t_n=\min\{k>t_{n-1}: d_k>0\}$.  Let
$b_n=d_{t_n}$. If $\liminf b_n = \infty$, then this test procedure is
consistent in the sense of Definition~\ref{consistency} over this
sequence of latent positions.
%given in Definition \eqref{consistency}, i 
%namely, for any
%$\beta > 0$ there exists a $n_2(\alpha, \beta, C)$ such that $\inf_{n
%  \geq n_2} \mathbb{P}(T_n \in R) \geq 1 - \beta$.  
\end{theorem}

\begin{figure}[htb!]
  \centering
\begin{tikzpicture}[cap=round,>=latex,every node/.style={scale=1}]
        % \draw[->] (-4.5cm,0cm) -- (7cm,cm) node[right,fill=white]{$x$};
        % \draw[->] (0cm,-4.5cm) -- (0cm,4.5cm) node[above,fill=white]{$y$};
        \def\a{4}
        \tkzInit[xmin=-.4,xmax=5,ymin=-.4,ymax=5] 
        \tkzDefPoint(0,0){O} \tkzDefPoint(\a,0){S1} \tkzDefPoint(0,\a){S2}
        \tkzDrawSector[color=blue](O,S1)(S2)
        \tkzDrawXY[noticks] 
       % \draw[thick] (0cm,0cm) circle(3cm);
% Source of rays
        \coordinate (s) at (0cm,0cm); % \node[above left] at (-3cm,0)
        % {$(-1,0)$}; 
        \path(s) ++({atan2(1,0.5)}:2) coordinate(p);
        \path(s) ++({atan2(1,1)}:2) coordinate(q);
        \path(s) ++({atan2(1,1)}:3.5) coordinate(r);
        \fill[red] (p) circle[radius=2pt] ++({atan2(1,1.25)}:1em)
        node{p};
        \fill[red] (q) circle[radius=2pt] ++({atan2(1, 0.25)}:1em) node {q};
        \fill[red] (r) circle[radius=2pt] ++({atan2(1,-0.25)}:1em) node {r};
        \draw[orange] (s)--++({atan2(1,0.5)}:4) node[black,right] {$\theta_{p}= \tfrac{1}{2}$}; 
        \draw[orange] (s)--++({atan2(1,1)}:4) node[black,right]
        {$\theta_{q}= \theta_{r}= 1 $}; 
\end{tikzpicture}
\caption{A pictorial example to illustrate the effect of
  projection. The distance between $p$ and $q$ is originally small,
  but increases after projection of $p$ to $\theta_{p} = 1/2$ and $q$
  to $\theta_{q} = 1$. The distance between $q$ and $r$ after
  projection is zero and the distance between $p$ and $r$ after
  projection decreases.}
\label{fig:projection}
\end{figure}
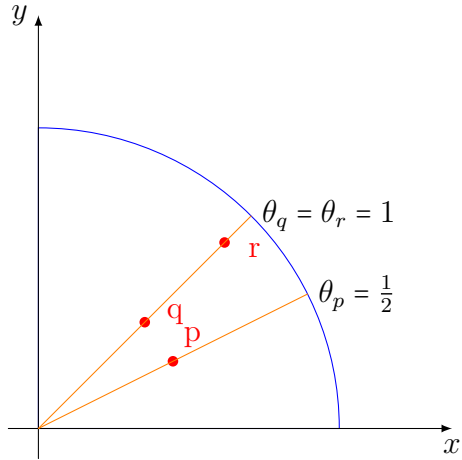\begin{remark}
  If the latent positions of $\mathbf{X}$ and $\mathbf{Y}$ are related
  by a diagonal transformation, this implies that each row $X_i$ of
  $\mathbf{X}$ is a scaled version of the corresponding row $Y_i$ of
  $\mathbf{Y}$; that is, $X_i=c_i Y_i$.  Under the null, the angle
  between the adjacency spectral embeddings $\hat{X}_i$ and
  $\hat{Y}_i$ should be small.  This suggests that we consider a
  cosine distance between the rows, and the projection in the
  numerator of our test statistic is essentially just that: namely, it
  measures the distance between projections of rows of the latent
  positions on the sphere (see
  Figure~\ref{fig:projection}). 
  % between true and estimated latent positions. 
  There are several other
  reasonable choices of test statistic; ours happens to be
  straightforward to analyze, and the denominator is a natural
  % consequence of a collection of known bounds on the accuracy of the
  upper bound on the numerator under the null hypothesis
  $\mathbf{X}_n \upVdash \mathbf{D}_n \mathbf{Y}_n$.
  Figure~\ref{fig:projection} also indicates that a latent position
  $X_i$ and its estimate $\hat{X}_i$ that are both in a sufficiently small
  $\epsilon$-neighborhood of the origin, and hence close, could have
  projections onto the sphere that are far apart. The lower bound
  condition on $\min_{i} \|X_i\|$ in
  Assumption~\ref{eigengap_assump_diagonal} addresses this 
  issue by requiring that the latent positions are not too ``small''
  compared to the density of the graph itself; that is, 
  ``small'' values of $\min_{\mathbf{W}}
  \|\hat{\mathbf{X}} \mathbf{W} - \mathbf{X} \|_{F}$ imply ``small''
  values of $\min_{\mathbf{W}} \|\mathcal{P}(\hat{\mathbf{X}}) \mathbf{W} -
  \mathcal{P}(\mathbf{X}) \|_{F}$ and similarly ``small'' values of
  $\min_{\mathbf{W}} \|\mathbf{X} \mathbf{W} - \mathbf{Y} \|_{F}$
  imply ``small'' values of $\min_{\mathbf{W}} \|\mathcal{P}(\mathbf{X}) \mathbf{W} -
  \mathcal{P}(\mathbf{Y}) \|_{F}$. 
\end{remark}

We illustrate the test procedure by a simulation example. 
% Next we consider the hypothesis test $H_0 \colon \mathbf{X}_n
% \upVdash \mathbf{D}_n \mathbf{Y}_n$ for some diagonal matrix
% $\mathbf{D}_n$ with positive diagonal entries against the alternative
% $\mathbf{X}_n \nupVdash \mathbf{D}_n \mathbf{Y}_n$ for all diagonal
% matrices $\mathbf{D}_n$ with positive diagonal entries; 
In particular,
we focus here on degree-corrected stochastic blockmodels
\citep{karrer2011stochastic} with block probability vector $\bm{\pi} =
(0.4,0.6)$ and block probability matrices $\mathbf{B}_0$,
$\mathbf{B}_1$ and $\mathbf{B}_2$ where
\begin{equation*}
  \mathbf{B}_0 = \begin{bmatrix} 0.5 & 0.2 \\ 0.2 &
    0.5 \end{bmatrix}; \quad \mathbf{B}_2 = \begin{bmatrix} 0.72 &
    0.192 \\ 0.192 & 0.32 \end{bmatrix} = \begin{bmatrix} 1.2 & 0 \\
    0 & 0.8 \end{bmatrix} \mathbf{B}_0 \begin{bmatrix} 1.2 & 0 \\ 0
    & 0.8 \end{bmatrix}; 
  \quad \mathbf{B}_1 
  = \begin{bmatrix} 0.7 & 0.2 \\ 0.2 & 0.7 \end{bmatrix}.
\end{equation*}
Recall that a degree corrected stochastic blockmodel graph $G$ on
$n$ vertices with $K$ blocks is parametrized by a block probability
vector $\pi \in \mathbb{R}^{K}$, a $K \times K$ block probability
matrix $\mathbf{B}$, and a degree correction vector $\bm{c} \in
\mathbb{R}^{n}$. The vertices of $G$ are assigned into one of the
$K$ blocks. The edges of $G$ are independent; furthermore, given
that vertices $i$ and $j$ are assigned into block $\tau(i)$ and
$\tau(j)$, the probability of an edge between $i$ and $j$ is simply
$c_i c_j \mathbf{B}_{\tau(i), \tau(j)}$. The vector $\bm{c}$ allows
for heterogeneity of degree within blocks, in contrast to the
homogeneity exhibited by traditional stochastic blockmodels.

\begin{figure}[htbp]
%    \captionsetup[subfig]{labelformat=empty}
    \centering
    \includegraphics[width=0.8\textwidth]{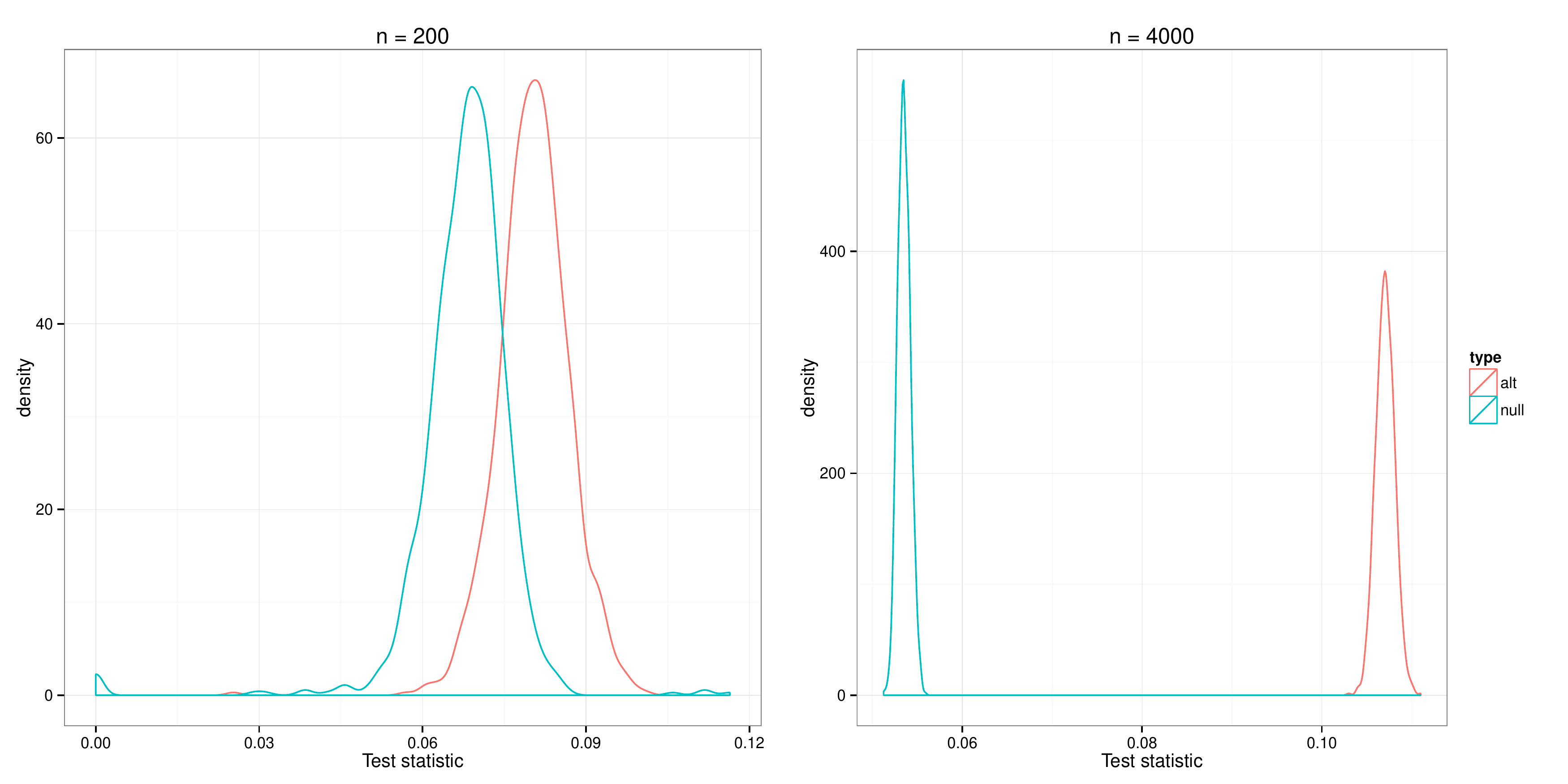}
%    \subfloat{
%    \includegraphics[width=0.8\textwidth]{test1_n1000.pdf}}
  \caption{Density estimate for the test statistic when testing $H_0
    \colon \mathbf{X}_n \upVdash \mathbf{D}_n \mathbf{Y}_n$ for some
    diagonal matrix $\mathbf{D}_n$
    against the alternative $H_{A} \colon \mathbf{X}_n \nupVdash \mathbf{D}_n
    \mathbf{Y}_n$ for all diagonal matrix $\mathbf{D}_n$.}
  \label{fig:testd}
  \end{figure}

By the above construction, $\mathbf{B}_2$ and $\mathbf{B}_0$
correspond to the same degree corrected stochastic blockmodel. We also
generate for each graph a vector of degree correction factors for the
vertices; these correspond to i.i.d. draws from a uniform
distribution on the interval $[0.2,1]$. The results are
presented in Figure~\ref{fig:testd} for $n = 200$ and $n =
4000$. The test once again exhibits good power when using the
rejection region obtained via the bootstrapping procedure.  

\section{Discussion}
\label{sec:discussion}
In summary, we show in this paper that the adjacency spectral
embedding can be used to generate simple and intuitive test statistics
for the inference problem of testing whether two
random dot product graphs on the same vertex set have the same or
related generating latent positions.  Two-sample graph inference has
significant applications in diverse fields; our test is both a
principled and, as our real data examples illustrate, practically
viable inference procedure. %  We expect that our procedure can also be
% extended to test similar hypotheses for more general latent position
% random graphs.

Our concentration inequalities allow us to obtain an at most
level-$\alpha$ consistent test without specifying the finite-sample or
asymptotic distribution of our test statistic.  We do not, at present,
have a limiting distributional result for our test statistic, and we
suspect that such a result would require additional, more restrictive,
model assumptions.

The test statistic based on orthogonal Procrustes matching
$\min_{\mathbf{W} \in \mathcal{O}(d)} \|\hat{\mathbf{X}} \mathbf{W} -
\hat{\mathbf{Y}} \|_{F}$ is but one of many possible test statistics
for testing the hypothesis $\mathbf{X} \upVdash \mathbf{Y}$. For
example, the test statistic $\|\mathbf{A} - \mathbf{B}\|_{F}$ is
intuitively appealing; it is a surrogate measure for the difference
$\|\mathbf{XX}^{T} - \mathbf{YY}^{T}\|_{F}$. Furthermore,
$\|\mathbf{A} - \mathbf{B}\|_{F}^{2} = 2 \sum_{i < j} (\mathbf{A}_{ij}
- \mathbf{B}_{ij})^2$ is a sum of independent Bernoulli random
variables; hence it is easily analyzable and may possibly yield more
powerful test. However, since $(\mathbf{A}_{ij} - \mathbf{B}_{ij})^2$
is a Bernoulli random variable with parameter $\mathbf{P}_{ij}(1 -
\mathbf{Q}_{ij}) + (1 - \mathbf{P}_{ij})\mathbf{Q}_{ij}$, this forces
that $(\mathbf{A}_{ij} - \mathbf{B}_{ij})^{2} \sim
\mathrm{Bernoulli}(1/2)$ if $\mathbf{Q}_{ij} = 1/2$, regardless of the
value of $\mathbf{P}_{ij}$. Therefore, $\|\mathbf{A} -
\mathbf{B}\|_{F}^{2} \sim \mathrm{Binomial}(\tbinom{n}{2},1/2)$
whenever $\mathbf{Q} = 1/2 \mathbf{J}$ where $\mathbf{J}$ is the
matrix of all ones. Thus, $\|\mathbf{A} - \mathbf{B}\|_{F}$ yields a
test that is not consistent for a large class of alternatives.

Yet another simple test statistic is based on the spectral norm
difference $\|\mathbf{A} - \mathbf{B}\|$; this is once again a
surrogate measure for the difference $\|\mathbf{XX}^{T} -
\mathbf{YY}^{T}\|$, and such a test statistic may be
more robust to model misspecification, e.g. when $\mathbf{A}$ and
$\mathbf{B}$ are adjacency matrices of more general latent position
random graphs. The concentration bound of
\cite{oliveira2009concentration}, which we state in
Eq.~\eqref{PA_diff} in Proposition~\ref{Old_bounds}, can be used to
construct a level-$\alpha$ test for the hypothesis $\mathbf{X}
\upVdash \mathbf{Y}$. However, the rejection region will be quite conservative and thus negatively impacts finite-sample performance. 
Thus, the development of a simple and principled way to bootstrap the test procedure in this context is an open question of some importance. 
Indeed, procedures for bootstrapping graphs and their statistics is currently a nascent field of research. 
See e.g, \citet{bickel:_subsampling} and Chapter 5 of \citet{kolaczyk:_statistical} for discussion of sampling procedures related to 
counting features in a network. Finally, we believe that test
statistics based directly on the adjacency matrices are also less
flexible. For instance, it is not obvious to us that such test
statistics can be easily adapted to test the hypothesis $\mathbf{X}
\upVdash \mathbf{D} \mathbf{Y}$ for some diagonal matrix $\mathbf{D}$,
or to conduct the nonparametric test of equality of the underlying
distributions for the latent positions a la \citet{tang14:_two:nonparametric}. 

%for example with $\mathbf{P}=\frac12\mathbf{J}$, and $\mathbf{Q}$ is arbitrary; 
%its distribution does not depend on $\mathbf{P}$ whenever $\mathbf{B}
%\sim \mathrm{Bernoulli}(1/2 \mathbf{J})$. 

To relate our test to classical generalized likelihood ratio tests, we
note that if we have two independent random dot product graphs with no
rank restrictions, the generalized likelihood ratio test statistic
reduces to
$$\Lambda=||\mathbf{A}-\mathbf{B}||_F^2$$
which is the aforementioned Frobenius norm test statistic.  However,
computing the generalized likelihood ratio test statistic under rank
assumptions is computationally more challenging.  We can approximate
this quantity by
\begin{equation*}
 \hat{\Lambda}=\frac{\mathbb{P}(\mathbf{A}|\mathbf{\hat{X}\hat{X}^T}) 
   \mathbb{P}(\mathbf{B}|\mathbf{\hat{X}\hat{X}^T})}{\mathbb{P}(\mathbf{A}|\mathbf{\hat{Z}})\mathbb{P}(\mathbf{B}|\mathbf{\hat{Z}})}
\end{equation*}
where
$\hat{\mathbf{Z}}=\frac{\mathbf{\hat{X}\hat{X}^T}+\mathbf{\hat{Y}\hat{Y}^T}}{2}$.
The question of how valid this approximation is, and how the limiting
distribution of this test statistic is related to ours, is the subject
of further research.  We emphasize that the likelihood ratio has an
independence assumption that we do not require. Also, since
$\hat{\mathbf{X}}$ is a consistent estimate for $\mathbf{X}$, our test statistic, which is a scaled version of $||\mathbf{X}-\mathbf{YW}||_F$, is in the spirit of a Wald test.
% Since both $\hat{\mathbf{X}}$ and the MLE for $\mathbf{X}$ are
% consistent in the appropriate norm, we surmise that our test
% statistic is close, if not equivalent, to a Wald test.

% This may, however, lead to a test that is less powerful for finite sample. 
% consistent for a narrower class of alternatives than that given in
% Theorem~\ref{thm:identity}. Indeed, by Eq.~\eqref{PA_diff}, the test
% $\|\mathbf{A} - \mathbf{B}\|$ is consistent when the sequence
% $((\delta(\mathbf{P}_n) + \delta(\mathbf{Q}_n)) \log{n})^{-1/2}
% \|\mathbf{P}_n - \mathbf{Q}_n\|$ diverges. In addition, we have
% \begin{equation*}
%   \begin{split}
%   \|\mathbf{P}_n - \mathbf{Q}_n\| &= \|\mathbf{X}_n \mathbf{X}_n^{T} -
%   \mathbf{Y}_n \mathbf{Y}_n^{T}\| \\ &= \|(\mathbf{X}_n \mathbf{W}_n -
%   \mathbf{Y}_n)(\mathbf{X}_n \mathbf{W}_n + \mathbf{Y}_n)^{T} + (\mathbf{X}_n \mathbf{W}_n +
%   \mathbf{Y}_n)(\mathbf{X}_n \mathbf{W}_n - \mathbf{Y}_n)^{T}\| \\
%   & \leq 2 \|\mathbf{X}_n \mathbf{W}_n - \mathbf{Y}_n\| \|\mathbf{X}_n
%   \mathbf{W}_n + \mathbf{Y}_n \| \leq 2 \|\mathbf{X}_n \mathbf{W}_n - \mathbf{Y}_n \|
%   \sqrt{\delta(\mathbf{P}_n) + \delta(\mathbf{Q}_n)}. 
%   \end{split}
% \end{equation*}
% for all orthogonal $\mathbf{W}_n$. Therefore, the fact that
% $(\delta(\mathbf{P}_n) + \delta(\mathbf{Q}_n))^{-1/2} \|\mathbf{P}_n -
% \mathbf{Q}_n\|$ diverges implies that $\min_{\mathbf{W} \in
%   \mathcal{O}(d)} \|\mathbf{X}_n \mathbf{W}_n - \mathbf{Y}_n\|$
% diverges, but not conversely, and hence $\|\mathbf{A} - \mathbf{B}\|$
% may be consistent for a narrower class of alternatives. 

Test statistics based on the spectral decomposition of the normalized
Laplacian matrices can also be constructed. However, the resulting
embedding is an estimate of some transformation of the latent
positions rather than the latent positions themselves. More
specifically, denote by $\tilde{\mathbf{X}}_n$ and
$\tilde{\mathbf{Y}}_n$ the spectral decomposition obtained from the
normalized Laplacian matrices associated with $\mathbf{A}_n$ and
$\mathbf{B}_n$, respectively. Then $\tilde{\mathbf{X}}_n$ is, up to
some orthogonal transformation, ``close'' to
$\mathcal{L}(\mathbf{X}_n)$ where $\mathcal{L}(\mathbf{X}_n)$ is a
transformation of $\mathbf{X}_n$, i.e., the $i$-th row of
$\mathcal{L}(\mathbf{X}_n)$ is given by $X_i/\langle X_i, \sum_{j \not
  = i} X_j \rangle$; similarly, $\tilde{\mathbf{Y}}_n$ is ``close'' to
$\mathcal{L}(\mathbf{Y}_{n})$ \citep[\S~6.3]{sussman12:_univer}. The
construction of test statistics for testing the hypothesis in
Section~\ref{sec:setting} for $\mathbf{X}_n$ and $\mathbf{Y}_n$ based
on the estimates $\tilde{\mathbf{X}}_n$ and $\tilde{\mathbf{Y}}_n$ of
$\mathcal{L}(\mathbf{X}_n)$ and $\mathcal{L}(\mathbf{Y}_n)$ is
certainly possible; however, subtle technical issues regarding
assumptions on the sequence of latent positions and speed of
convergence of the estimates $\tilde{\mathbf{X}}_n$ and
$\tilde{\mathbf{Y}}_n$ can arise. In summary, the formulation of the
hypotheses and the accompanying test procedures in
Section~\ref{sec:setting} are such that the test statistics are simple
functions of the adjacency spectral embeddings of the graphs. Other
formulations of comparable two-sample tests could, of course, lead to
test statistics that are simple functions of the normalized Laplacian
embeddings.  \bibliography{../biblio.bib}

\newpage
\appendix
\section{Additional lemmas and proofs}
\label{sec:additional-proofs}
\subsection*{Established bounds}
We first state a bound on the spectral norm difference between
$\mathbf{A}_n$ and $\mathbf{P}_n$. The bound is from Theorem 3.1 of \cite{oliveira2009concentration}.
%sussman12:_univer, tangs.:_univer, athreya2013limit, bickel_sarkar_2013}.
\begin{proposition}
  \label{Old_bounds}
  Let $\hat{{\bf X}}_n \in \mathbb{R}^{n \times d}$ be the adjacency
  spectral embedding of the $n \times n$ adjacency matrix ${\bf A}_n
  \sim \mathrm{Bernoulli}({\bf P}_n)$ where ${\bf P}_n= \mathbf{X}_n
  \mathbf{X}_n^{T}$ is of rank $d$ and its non-zero eigenvalues are
  distinct.  Suppose also that there exists $\epsilon>0$ such that
  $\delta(\mathbf{P}_n) \geq (\log n )^{1+ \epsilon}$. Let $c>0$ be
  arbitrary but fixed. There exists $n_0(c)$ such that if $n>n_0$ and
  $\eta$ satisfies $n^{-c} < \eta< 1/2$, then with probability at
  least $1-2\eta$, the following hold simultaneously. 
  \begin{equation}
    \label{PA_diff}
    \|{\bf P}_n-{\bf A}_n\| \leq 2 \sqrt{\delta(\mathbf{P}_n) \log (n/\eta)}
  \end{equation}
 \end{proposition}
Next, we state a simple proposition on the consistency of adjacency-based estimates of $\delta(\mathbf{P}_n), \gamma_1(\mathbf{P}_n),$ and $\gamma_2(\mathbf{P}_n)$.  This proposition is a straightforward consequence of Hoeffding's equality, Equation \eqref{PA_diff}, and the Borel-Cantelli Lemma, and we omit the proof.
%   and 
%   \begin{equation}
%     \label{spec_norm_oldbound}
%     \frac{{\bf S}_P(d,d)}{2} \leq \|{\bf S}_P\| \textrm{and } 
%     \frac{{\bf S}_P(d,d) }{2} \leq \|{\bf S}_A\|
% \end{equation}
%where $||\cdot ||_{2 \to 2}$ denotes the $2$-to-$2$ operator norm of a matrix.\\
%Finally, with probability at least $1-2\eta$,
%\begin{equation}
%\label{up_ua_close_to_id}
%\|{\bf U}_{\mathbf{P}_n}^T {\bf U}_{\mathbf{A}_n}-\mathbf{I}\|_F \leq 
%\frac{10 d \log (n/\eta)}{\gamma_1^2(\mathbf{P}_n) \delta(\mathbf{P}_n)}
%\end{equation}

\begin{proposition}
  \label{prop:gamma(A)}
  Let $\{\mathbf{X}_n\}$ be a sequence of latent positions and suppose
  that the sequence of matrices $\{{\bf P}_n\}$, where ${\bf
    P}_n=\mathbf{X}_n \mathbf{X}_n^{T}$, satisfy the condition in Eq.\eqref{eq:eigengap_delta} in
  Assumption~\ref{eigengap_assump}. Let $\{\mathbf{A}_n\}$ be the
  sequence of adjacency matrices $\mathbf{A}_n \sim
  \mathrm{Bernoulli}({\bf P}_n)$. Then we have
  \begin{equation}
    \label{eq:10}
    \frac{\delta(\mathbf{A}_n)}{\delta(\mathbf{P}_n)} 
    \overset{\mathrm{a.s.}}{\longrightarrow} 1; \quad 
    \frac{\gamma_1(\mathbf{A}_n)}{\gamma_{1}({\bf P}_n)} 
    \overset{\mathrm{a.s.}}{\longrightarrow} 1; \quad
    \frac{\gamma_2(\mathbf{A}_n)}{\gamma_{2}(\mathbf{P}_n)} 
    \overset{\mathrm{a.s.}}{\longrightarrow} 1; 
  \end{equation}
\end{proposition}

\subsection*{Additional lemmas}
Now, let $\mathbf{W}$ be such that $\mathbf{U}_{\mathbf{P}}
\mathbf{S}_{\mathbf{P}}^{1/2} = \mathbf{X} \mathbf{W}$. We note that
such a matrix $\mathbf{W}$ always exists as $\mathbf{U}_{\mathbf{P}}
\mathbf{S}_{\mathbf{P}} \mathbf{U}_{\mathbf{P}}^{T} = \mathbf{P} =
\mathbf{X} \mathbf{X}^{T}$. The proof of Theorem~\ref{thm:conc_Xhat_X}
proceeds by bounding, in a series of technical lemmas, each of the
terms in parentheses in the following decomposition of
$\hat{\mathbf{X}} - \mathbf{X} \mathbf{W}$:
\begin{equation*}
  \begin{split}
   \hat{\mathbf{X}} - \mathbf{X} \mathbf{W} &= \mathbf{U}_{\mathbf{A}}
  \mathbf{S}_{\mathbf{A}}^{1/2}  - \mathbf{U}_{\mathbf{P}}
  \mathbf{S}_{\mathbf{P}}^{1/2}  = \mathbf{A}
  \mathbf{U}_{\mathbf{A}} \mathbf{S}_{\mathbf{A}}^{-1/2} - 
  \mathbf{P} \mathbf{U}_{\mathbf{P}} \mathbf{S}_{\mathbf{P}}^{-1/2} \\
  & = \mathbf{A} (\mathbf{U}_{\mathbf{A}} - \mathbf{U}_{\mathbf{P}})
  \mathbf{S}_{\mathbf{A}}^{-1/2} + \mathbf{A} \mathbf{U}_{\mathbf{P}} (
  \mathbf{S}_{\mathbf{A}}^{-1/2} - \mathbf{S}_{\mathbf{P}}^{-1/2}) +
  (\mathbf{A} - \mathbf{P})  \mathbf{U}_{\mathbf{P}}
  \mathbf{S}_{\mathbf{P}}^{-1/2}
  \end{split}
\end{equation*}
We now state these lemmas, beginning with two results: the first is
Lemma 10 of \cite{lyzinski13:_perfec}, and it provides a bound for
$\|(\mathbf{A} \mathbf{U}_{\mathbf{A}} \mathbf{S}_{\mathbf{A}}^{-1/2}
- \mathbf{A} \mathbf{U}_{\mathbf{P}}
\mathbf{S}_{\mathbf{A}}^{-1/2}\|_{F}$ by viewing it as the difference
after one step of the power method for $\mathbf{A}$ when starting at
$\mathbf{U}_{\mathbf{P}}$. The second bounds
$\|\mathbf{A}\mathbf{U}_{\mathbf{P}}(\mathbf{S}_{\mathbf{A}}^{-1/2} -
\mathbf{S}_{\mathbf{P}}^{-1/2})\|_F$ using Lemma 2 of
\cite{athreya2013limit} and the expansion
$$\mathbf{S}_{\mathbf{A}}^{-1/2} -\mathbf{S}_{\mathbf{P}}^{-1/2}=
(\mathbf{S}_{\mathbf{P}}-\mathbf{S}_{\mathbf{A}})
(\mathbf{S}_{\mathbf{P}}^{1/2}+ \mathbf{S}_{\mathbf{A}}^{1/2})^{-1}
(\mathbf{S}_{\mathbf{A}}^{-1/2}\mathbf{S}_{\mathbf{P}}^{-1/2})$$  

\begin{lemma}
  \label{lem:1}
  If the events in Proposition~\ref{Old_bounds} occur, then 
  \begin{equation}
    \label{eq:3}
    \|\mathbf{A} \mathbf{U}_{\mathbf{A}}
    \mathbf{S}_{\mathbf{A}}^{-1/2}  - \mathbf{A} \mathbf{U}_{\mathbf{P}}
    \mathbf{S}_{\mathbf{A}}^{-1/2} \|_{F} \leq \frac{24 \sqrt{2} d
      \log{(n/\eta)}}{\sqrt{\gamma_{1}^{5}(\mathbf{P}) \delta(\mathbf{P})}}
  \end{equation}
\end{lemma}

\begin{lemma}
  \label{lem:3}
   If the events in Proposition~\ref{Old_bounds} occur, then 
   \begin{equation}
     \label{eq:6}
   \|\mathbf{A}\mathbf{U}_{\mathbf{P}}(\mathbf{S}_{\mathbf{A}}^{-1/2} -
\mathbf{S}_{\mathbf{P}}^{-1/2})\|_F \leq 
    \frac{18 d^{3/2} \log{(n/\eta)}}{\sqrt{\gamma_{1}^{7}(\mathbf{P}) \delta(\mathbf{P})}}.
   \end{equation}
\end{lemma}

Our last technical lemma is a concentration
bound for $\|(\mathbf{A} - \mathbf{P}) \mathbf{U}_{\mathbf{P}}
\mathbf{S}_{\mathbf{P}}^{-1/2}\|_{F}$ whose proof is given in the
following subsection. 
\begin{lemma}
  \label{lem:4}
  Let $\eta > 0$ be arbitrary. Then with probability at least $1 -
  2\eta$, the events in Proposition~\ref{Old_bounds} occur and
  furthermore, 
  \begin{equation}
    \label{eq:7}
    \bigl|\|(\mathbf{A} - \mathbf{P}) \mathbf{U}_{\mathbf{P}}
\mathbf{S}_{\mathbf{P}}^{-1/2} \|^{2}_{F} - C^{2}(\mathbf{X}) \bigr| \leq
\frac{14 \sqrt{2d} \log{(n/\eta)}}{\gamma_{2}(\mathbf{P}) \sqrt{\delta(\mathbf{P})}}.
  \end{equation}
  where $C^{2}(\mathbf{X})$ is the following function of $\mathbf{X}$:
  \begin{equation*}
    C^{2}(\mathbf{X}) = \mathrm{tr} \,\, 
    \mathbf{S}_{\mathbf{P}}^{-1/2} \mathbf{U}_{\mathbf{P}}^{T} \mathbb{E}[(\mathbf{A} -
    \mathbf{P})^{2}] \mathbf{U}_{\mathbf{P}}
    \mathbf{S}_{\mathbf{P}}^{-1/2} = \mathrm{tr} \,\,
    \mathbf{S}_{\mathbf{P}}^{-1/2} \mathbf{U}_{\mathbf{P}}^{T} \mathbf{D}
    \mathbf{U}_{\mathbf{P}} \mathbf{S}_{\mathbf{P}}^{-1/2} \leq
    d\gamma_{2}^{-1}(\mathbf{P})
  \end{equation*}
  and $\mathbf{D}$ is a diagonal matrix whose diagonal entries are
  given by
  \begin{equation*}
    \mathbf{D}_{ii} = \sum_{k \not = i} \mathbf{P}_{ik} (1 - \mathbf{P}_{ik}).
  \end{equation*}
\end{lemma}

\subsection*{Proofs of main results}
We now provide proofs of the main results in the paper, starting with Lemma \ref{lem:4}.

{\bf Proof of Lemma \ref{lem:4}}
Let $Z = \|(\mathbf{A} - \mathbf{P}) \mathbf{U}_{\mathbf{P}}
\mathbf{S}_{\mathbf{P}}^{-1/2} \|^{2}_{F}$. Since our graphs are
undirected and loop free, $Z$ is a function of the $n(n-1)/2$
independent random variables $\{\mathbf{A}_{ij}\}_{i < j}$. Let
${\bf A}$ and $\mathbf{A}'$ be two arbitrary adjacency
matrices. Denote by $\mathbf{A}^{(kl)}$ the adjacency matrix
obtained by replacing the $(k,l)$ and $(l,k)$ entries of
$\mathbf{A}$ by those of $\mathbf{A}'$. Let $Z_{kl} =
\|(\mathbf{A}^{(kl)} - \mathbf{P}) \mathbf{U}_{\mathbf{P}}
\mathbf{S}_{\mathbf{P}}^{-1/2} \|_{F}^{2}$. The
argument we employ is based on the following logarithmic Sobolev concentration
inequality for $Z - \mathbb{E}[Z]$ \cite[\S 6.4]{boucheron13:_concen_inequal}. 
\begin{theorem}
  \label{thm:log_Sobolev}
  Assume that there exists a constant $v > 0$ such that, with
  probability at least $1 - \eta$,
  \begin{equation*}
    \sum_{k < l} (Z - Z_{kl})^{2} \leq v.
  \end{equation*}
  Then for all $t > 0$, 
  \begin{equation*}
    \mathbb{P}[|Z - \mathbb{E}[Z]| > t] \leq 2e^{-t^2/(2v)} + \eta.
  \end{equation*}
\end{theorem}
Let $\mathbf{V} = \mathbf{U}_{\mathbf{P}}
\mathbf{S}_{\mathbf{P}}^{-1/2}$. For notational convenience, we denote
the $i$-th row of $\mathbf{V}$ by $V_i$. We shall also denote the
inner product between vectors in Euclidean space by $\langle \cdot,
\cdot \rangle$. The $i$-th row of the product
$(\mathbf{A} - \mathbf{P}) \mathbf{V}$ is simply a linear combination
of the rows of $\mathbf{V}$, i.e.,
\begin{equation*}
  ((\mathbf{A} - \mathbf{P}) \mathbf{V})_{i} = \sum_{j=1}^{n}
  (\mathbf{A} - \mathbf{P})_{ij} V_j.
\end{equation*}
Hence,
\begin{equation*}
  \begin{split}
  Z &= \|(\mathbf{A} - \mathbf{P}) \mathbf{V} \|_{F}^{2} 
  = \sum_{i=1}^{n} \|((\mathbf{A} - \mathbf{P}) \mathbf{V})_{i}
  \|^{2}
  = \sum_{i=1}^{n} \sum_{j=1}^{n} \sum_{k=1}^{n} (\mathbf{A} -
  \mathbf{P})_{ij} (\mathbf{A} - \mathbf{P})_{ik} \langle V_{j}, V_{k} \rangle
  \end{split}
\end{equation*}
As $\mathbf{A}$ and $\mathbf{A}^{(kl)}$ differs possibly only in the $(k,l)$
and $(l,k)$ entries and that the entries of $\mathbf{A}$ and
$\mathbf{A'}$ are binary variables, we have that if $(Z - Z_{kl})$ is non-zero, then
\begin{equation*}
  \begin{split}
  Z - Z_{kl} &= 2 \Bigl(\sum_{j \not = l} (\mathbf{A} -
  \mathbf{P})_{kj} \langle V_j, V_l \rangle \Bigr) + 2 \Bigl( \sum_{j
    \not = k} (\mathbf{A} - \mathbf{P})_{lj} \langle V_j, V_k \rangle \Bigr)
  + (1 - 2 \mathbf{P}_{kl}) \langle V_{l}, V_k \rangle \\
  &= 2 \sum_{j=1}^{n} (\mathbf{A} -
  \mathbf{P})_{kj} \langle V_j, V_l \rangle \Bigr) + 2 \sum_{j=1}^{n}
  (\mathbf{A} - \mathbf{P})_{lj} \langle V_j, V_k \rangle \Bigr) +
  c_{kl}
  \end{split}
\end{equation*}
where $c_{kl} = 2(\mathbf{A} - \mathbf{P})_{kl} \langle V_{l}, V_{l} \rangle + 2
  (\mathbf{A} - \mathbf{P})_{lk} \langle V_{k}, V_{k} \rangle + (1 - 2
  \mathbf{P})_{kl} \langle V_l, V_k \rangle$. We then have
  \begin{equation*}
    (Z - Z_{kl})^{2} \leq 3(C_{kl}^{(1)} + C_{kl}^{(2)} + c_{kl}^{2})
  \end{equation*}
where $C_{kl}^{(1)}$ and $C_{kl}^{(2)}$ are given by
\begin{gather*}
  C_{kl}^{(1)} = 4 \sum_{j_1 = 1}^{n} \sum_{j_2 = 1}^{n}(\mathbf{A} -
  \mathbf{P})_{kj_1} (\mathbf{A} - \mathbf{P})_{k j_2} \langle V_l,
  V_{j_1} \rangle \langle V_l, V_{j_2} \rangle = 4 \Bigl[((\mathbf{A} -
  \mathbf{P})\mathbf{V} \mathbf{V}^{T})_{kl}\Bigr]^{2} \\
  C_{kl}^{(2)} = 4 \sum_{j_{1} = 1}^{n} \sum_{j_{2} = 1}^{n} (\mathbf{A} -
  \mathbf{P})_{lj_1} (\mathbf{A} - \mathbf{P})_{l j_2} \langle V_k,
  V_{j_1} \rangle \langle V_k, V_{j_2} \rangle = 4 \Bigl[((\mathbf{A} -
  \mathbf{P})\mathbf{V} \mathbf{V}^{T})_{lk}\Bigr]^{2} 
\end{gather*}
As $C_{kl}^{(1)} = C_{lk}^{(2)}$, $c_{kl} = c_{lk}$, and $C_{kk}^{(1)} > 0$ for all $l,k$, we thus have
\begin{equation*}
  \begin{split}
  \sum_{k < l} (Z - Z_{kl})^{2} & \leq 3
  \sum_{k < l} \Bigl(C_{kl}^{(1)} + C_{kl}^{(2)} + c_{kl}^{2}\Bigr)
%   \leq 3 \sum_{k \not = l} C_{kl}^{(1)} + \frac{3}{2} \sum_{k \not =
%    l} c_{kl}^{2} 
  \leq 3 \sum_{k=1}^{n} \sum_{l=1}^{n} C_{kl}^{(1)} + \frac{3}{2}
  \sum_{k=1}^{n} \sum_{l=1}^{n} c_{kl}^{2}
  \end{split}
\end{equation*}
We now consider each of the term in the above right hand side.
\begin{equation*}
  \begin{split}
  \sum_{k=1}^{n} \sum_{l=1}^{n} C_{kl}^{(1)} &= %
4 \sum_{k=1}^{n} \sum_{l=1}^{n} \Bigl[((\mathbf{A} -
  \mathbf{P})\mathbf{V} \mathbf{V}^{T})_{kl}\Bigr]^{2} 
%   \sum_{l=1}^{n} \sum_{j_1=1}^{n} \sum_{j_2=1}^{n} (\mathbf{A} -
%   \mathbf{P})_{k j_1} (\mathbf{A} - \mathbf{P})_{k j_2} \langle V_{l},
%   V_{j_1} \rangle \langle V_{l}, V_{j_2} \rangle \\
% &= \sum_{j_1 =1}^{n} \sum_{j_2=1}^{n} \sum_{l=1}^{n} \sum_{k=1}^{n} (\mathbf{A} - \mathbf{P})_{k j_1}
% (\mathbf{A} - \mathbf{P})_{k j_2} \langle V_{l}, V_{j_1} \rangle
% \langle V_{l}, V_{j_2} \rangle \\
% &= \sum_{j_1=1}^{n} \sum_{j_2=1}^{n} \sum_{l=1}^{n} \sum_{k=1}^{n}
% (\mathbf{A} - \mathbf{P})_{k j_1} (\mathbf{A} - \mathbf{P})_{j_2 k} \langle V_{l}, V_{j_1} \rangle
% \langle V_{l}, V_{j_2} \rangle \\ 
% &= \sum_{j_1=1}^{n} \sum_{j_2=1}^{n} \sum_{l=1}^{n} ((\mathbf{A} -
% \mathbf{P})^{2})_{j_1 j_2} \langle V_{l}, V_{j_1} \rangle
% \langle V_{l}, V_{j_2} \rangle \\
% &= \sum_{j_1=1}^{n} \sum_{j_2=1}^{n} \sum_{l=1}^{n} ((\mathbf{A} -
% \mathbf{P})^{2})_{j_1 j_2} (\mathbf{V} \mathbf{V}^{T})_{j_1 l}
% (\mathbf{V} \mathbf{V}^{T})_{l j_2} \\
% &= \sum_{j_1=1}^{n} \sum_{j_2=1}^{n} ((\mathbf{A} -
% \mathbf{P})^{2})_{j_1 j_2} (\mathbf{V} \mathbf{V}^{T} \mathbf{V}
% \mathbf{V}^{T})_{j_1 j_2} \\
= 4 \|(\mathbf{A} - \mathbf{P}) \mathbf{V}
\mathbf{V}^{T} \|_{F}^{2}
  \end{split}
\end{equation*}
% \begin{equation*}
%   \begin{split}
%   \sum_{k=1}^{n} C_{kk}^{(1)} &= \sum_{k=1}^{n} \sum_{l=1}^{n}
%   (\mathbf{A} - \mathbf{P})_{k j_1} (\mathbf{A} - \mathbf{P})_{k j_2}
%   \langle V_{k}, V_{j_1} \rangle \langle V_{k}, V_{j_2} \rangle \\
%   &= \sum_{k=1}^{n} \sum_{j_1=1}^{n} \sum_{j_2=1}^{n} (\mathbf{A} - \mathbf{P})_{k j_1} (\mathbf{V}
%   \mathbf{V}^{T})_{j_1 k} (\mathbf{A} - \mathbf{P})_{k j_2}
%   (\mathbf{V} \mathbf{V}^{T})_{j_2 k} \\
%   &= \sum_{k=1}^{n} \sum_{j_1=1}^{n} (\mathbf{A} - \mathbf{P})_{k j_1} (\mathbf{V}
%   \mathbf{V}^{T})_{j_1 k}  ((\mathbf{A} - \mathbf{P})\mathbf{V}
%   \mathbf{V}^{T})_{kk} \\
%   &= \sum_{k=1}^{n} ((\mathbf{A} - \mathbf{P})\mathbf{V}
%   \mathbf{V}^{T})_{kk}^{2} = \|\mathrm{diag}((\mathbf{A} - \mathbf{P})
%   \mathbf{V} \mathbf{V}^{T}) \|_{F}^{2}
%  \end{split}
% \end{equation*}
\begin{equation*}
  \begin{split}
  \sum_{k=1}^{n} \sum_{l=1}^{n} c_{kl}^{2} &\leq 3 \sum_{k=1}^{n}
  \sum_{l=1}^{n} 4 (\mathbf{A} - \mathbf{P})_{kl}^{2} (\langle V_{l},
  V_{l} \rangle^{2} + \langle V_{k}, V_{k} \rangle^{2}) +
  3 \sum_{k=1}^{n} \sum_{l=1}^{n} \langle V_{l}, V_{k} \rangle^{2} \\ 
  &= 6 \sum_{k=1}^{n} 4((\mathbf{A} - \mathbf{P})^{2})_{kk} \langle
  V_{k}, V_{k} \rangle^{2} +  
  3 \sum_{k=1}^{n} \sum_{l=1}^{n} \langle V_{l}, V_{k} \rangle^{2} \\
  &= 24 \sum_{k=1}^{n} ((\mathbf{A} - \mathbf{P})^{2})_{kk} \langle
  V_{k}, V_{k} \rangle^{2} + 3 \sum_{k=1}^{n} (\mathbf{V}
  \mathbf{V}^{T} \mathbf{V} \mathbf{V}^{T})_{kk} \\ 
  &= 24 \sum_{k=1}^{n} ((\mathbf{A} - \mathbf{P})^{2})_{kk} \langle
  V_{k}, V_{k} \rangle^{2} + 3 \|\mathbf{V} \mathbf{V}^{T}\|_{F}^{2}
  \\
 & \leq 24  \|(\mathbf{A} - \mathbf{P})^{2}\| \sum_{k=1}^{n}
 \langle V_{k}, V_{k} \rangle^{2} + 3 \|\mathbf{V} \mathbf{V}^{T}\|_{F}^{2}
  \\ 
  & \leq 24 \|\mathbf{A} - \mathbf{P}\|^{2} \|\mathrm{diag}(\mathbf{V}
  \mathbf{V}^{T})\|_{F}^{2} + 3 \|\mathbf{V} \mathbf{V}^{T} \|_{F}^{2}
 \end{split}
\end{equation*}
where the penultimate inequality of the above display follows from the fact that the diagonal
elements of $(\mathbf{A} - \mathbf{P})^{2}$ is majorized by its
eigenvalues. % We then note that
% \begin{equation*}
%   \begin{split}
%    \|\mathrm{diag}(\mathbf{V} \mathbf{V}^{T}) \|_{F}^{2} &= \|
%   \mathrm{diag}(\mathbf{U}_{\mathbf{P}} \mathbf{S}_{\mathbf{P}}^{-1}
%   \mathbf{U}_{P}^{T}) \|_{F}^{2} \\ &=
%   \|\mathrm{diag}(\mathbf{U}_{\mathbf{P}}
%   \mathbf{S}_{\mathbf{P}}^{1/2} \mathbf{S}_{\mathbf{P}}^{-2}
%   \mathbf{S}_{\mathbf{P}}^{1/2} \mathbf{U}_{\mathbf{P}}^{T})
%   \|_{F}^{2} \\
%   & \leq \|\mathrm{diag}(\mathbf{U}_{\mathbf{P}}
%   \mathbf{S}_{\mathbf{P}}^{1/2} \mathbf{S}_{\mathbf{P}}^{1/2}
%   \mathbf{U}_{\mathbf{P}}^{T}) \|_{F}^{2} 
%   \|\mathbf{S}_{\mathbf{P}}^{-2} \|^{2} \\
%   &\leq n p^{2}_{\mathrm{max}} \frac{1}{(\gamma_{2} \delta)^{4}}
%   \end{split}
% \end{equation*}
% where 
We therefore have
\begin{equation*}
  \begin{split}
  \sum_{k < l}(Z - Z_{kl})^{2} &\leq \bigl(48 \|\mathbf{A} -
  \mathbf{P}\|^{2} + \tfrac{9}{2}\bigr) \|\mathbf{V}
  \mathbf{V}^{T} \|_{F}^{2} \\ &\leq 49 \|\mathbf{A} - \mathbf{P} \|^{2}
  \|\mathbf{V} \mathbf{V}^{T} \|_{F}^{2} \\ & = 49 \|\mathbf{A} -
  \mathbf{P} \|^{2} \|\mathbf{S}_{\mathbf{P}}^{-1} \|_{F}^{2} \\
  & \leq 49 \|\mathbf{A} - \mathbf{P}\|^{2} 
\frac{d}{(\gamma_{2}(\mathbf{P}) \delta(\mathbf{P}))^{2}}
  \end{split}
\end{equation*}
By Proposition~\ref{Old_bounds}, for any $\eta > 0$, with probability
at least $1 - \eta$,
\begin{equation*}
  \|\mathbf{A} - \mathbf{P}\|^{2} \leq 4 \delta(\mathbf{P}) \log{(n/\eta)}
\end{equation*}
Hence, for all $\eta > 0$, with probability at least $1 - \eta$, 
\begin{equation}
  \label{eq:2}
  \sum_{k < l} (Z - Z_{kl})^{2} \leq \frac{196 d
    \log{(n/\eta)}}{\gamma_{2}^{2}(\mathbf{P}) \delta(\mathbf{P})}.
\end{equation}
Denote by $v(\eta)$ the right hand side of the above display. 
We then have, by Theorem~\ref{thm:log_Sobolev}, that for all $t > 0$, 
\begin{equation}
  \label{eq:1}
  \mathbb{P}[|Z - \mathbb{E}[Z]| > t] \leq 2e^{-t^2/(2v(\eta))} + \eta
\end{equation}
Setting $t$ to be
\begin{equation*}
  t = \frac{14 \sqrt{2d} \log{(n/\eta)}}{\gamma_{2}(\mathbf{P}) \sqrt{\delta(\mathbf{P})}}
\end{equation*}
yields $2e^{-t^2/(2v(\eta))} \leq \eta$ as desired. 

Finally, we provide a bound for $\mathbb{E}[Z]$ in terms of the parameters
$\gamma_{2}(\mathbf{P})$. We have
\begin{equation*}
  \mathbb{E}[Z] = \mathbb{E}[\|(\mathbf{A} - \mathbf{P})\mathbf{V}
  \|_{F}^{2}] = 
\mathbb{E}[\mathrm{tr} \Bigl(
  \mathbf{V}^{T} (\mathbf{A} - \mathbf{P})^{2} \mathbf{V}
  \Bigr)] = \mathrm{tr} \Bigl(\mathbf{V}^{T}
  \mathbb{E}[(\mathbf{A} - \mathbf{P})^{2}] \mathbf{V} \Bigr)
\end{equation*}
We note that
\begin{equation*}
  \mathbb{E}[((\mathbf{A} - \mathbf{P})^{2})_{ij}] = \mathbb{E}\Bigl[
  \sum_{k} (\mathbf{A} - \mathbf{P})_{ik} (\mathbf{A} -
  \mathbf{P})_{kj}\Bigr] = \begin{cases} 0 & \text{if $i \not = j$} \\
    \sum_{k \not = i} \mathbf{P}_{ik} (1 - \mathbf{P})_{ik} & \text{if
      $i = j$}
    \end{cases}
\end{equation*}
Hence, $\delta(\mathbf{P}) \mathbf{I} - \mathbb{E}[(\mathbf{A} - \mathbf{P})^2] 
$ is positive semidefinite. We thus have
\begin{equation*}
  \mathbb{E}[\|(\mathbf{A} - \mathbf{P})\mathbf{V}
  \mathbf{V}^{T}\|_{F}^{2}] \leq \delta(\mathbf{P}) \mathrm{tr} 
  \mathbf{V}^{T} \mathbf{V} \leq d \gamma_{2}^{-1}(\mathbf{P}).
\end{equation*}
which establishes the upper bound $C^{2}(\mathbf{X}) \leq d
\gamma_{2}^{-1}(\mathbf{P})$ as required. 
\qed

{\bf Proof of Theorem~\ref{thm:conc_Xhat_X}.}
From Lemma~\ref{lem:4}, we have 
\begin{equation*}
  \bigl| \|(\mathbf{A} - \mathbf{P})\mathbf{U}_{\mathbf{P}}
    \mathbf{S}_{\mathbf{P}}^{-1/2}\|_{F}^{2} - C^{2}(\mathbf{X}) \bigr|
    \leq \frac{14\sqrt{2d} \log{(n/\eta)}}{\gamma_{2}(\mathbf{P}) \sqrt{\delta(\mathbf{P})}}
\end{equation*}
with probability at least $1 - 2\eta$. Now, $a \leq b + c$ implies
$\sqrt{a} \leq \sqrt{b} + \tfrac{c}{2\sqrt{b}}$ and $a \geq b
- c \geq 0$ implies $\sqrt{a} \geq \sqrt{b} - \tfrac{c}{\sqrt{b}}$. Hence
\begin{equation*}
  - \frac{14 \sqrt{2d} \log{(n/\eta)}}{C(\mathbf{X}) \gamma_{2}(\mathbf{P})
    \sqrt{\delta(\mathbf{P})}} \leq \|(\mathbf{A} - \mathbf{P}) \mathbf{U}_{\mathbf{P}}
  \mathbf{S}_{\mathbf{P}}^{-1/2} \|_{F} - C(\mathbf{X}) \leq \frac{7\sqrt{2d} \log{(n/\eta)}}{C(\mathbf{X}) \gamma_{2}(\mathbf{P}) \sqrt{\delta(\mathbf{P})}}
\end{equation*}
with probability at least $1 - 2\eta$. Applying Lemma~\ref{lem:1} and
Lemma~\ref{lem:3} yield 
\begin{equation*}
 - \frac{C_1 d^{3/2}\log{(n/\eta)}}{C(\mathbf{X}) \sqrt{\gamma_{1}^{7}(\mathbf{P}) \delta(\mathbf{P})}}
\leq \|\hat{\bf X}- \mathbf{P} \mathbf{U}_{\mathbf{P}}
\mathbf{S}_{\mathbf{P}}^{-1/2} \|_F-C({\bf X}) \leq \frac{C_2 d^{3/2}
\log{(n/\eta)}}{C(\mathbf{X}) \sqrt{\gamma_{1}^{7}(\mathbf{P}) \delta(\mathbf{P})}}
\end{equation*}
for some constants $C_1, C_2 > 0$. Finally, $\mathbf{P}
\mathbf{U}_{\mathbf{P}} \mathbf{S}_{\mathbf{P}}^{-1/2} =
\mathbf{X}
\mathbf{W}$ for some $\mathbf{W} \in \mathcal{O}(d).$ \qed
% Finally, we provide a bound for $\mathbb{E}[Z]$ given the parameters
% $\gamma_{2}$ and $\delta$. We have
% \begin{equation*}
%   \mathbb{E}[Z] = \mathbb{E}[\|(\mathbf{A} - \mathbf{P})\mathbf{V}
%   \|_{F}^{2}] = 
% \mathbb{E}[\mathrm{tr} \Bigl(
%   \mathbf{V}^{T} (\mathbf{A} - \mathbf{P})^{2} \mathbf{V}
%   \Bigr)] = \mathrm{tr} \Bigl(\mathbf{V}^{T}
%   \mathbb{E}[(\mathbf{A} - \mathbf{P})^{2}] \mathbf{V} \Bigr)
% \end{equation*}
% We note that
% \begin{equation*}
%   \mathbb{E}[((\mathbf{A} - \mathbf{P})^{2})_{ij}] = \mathbb{E}\Bigl[
%   \sum_{k} (\mathbf{A} - \mathbf{P})_{ik} (\mathbf{A} -
%   \mathbf{P})_{kj}\Bigr] = \begin{cases} 0 & \text{if $i \not = j$} \\
%     \sum_{k \not = i} \mathbf{P}_{ik} (1 - \mathbf{P})_{ik} & \text{if
%       $i = j$}
%     \end{cases}
% \end{equation*}
% Hence, $\mathbb{E}[(\mathbf{A} - \mathbf{P})^2] \prec \delta
% \mathbf{I}$ where $\prec$ refers to the partial ordering on the
% positive definite cone. We thus have
% \begin{equation*}
%   \mathbb{E}[\|(\mathbf{A} - \mathbf{P})\mathbf{V}
%   \mathbf{V}^{T}\|_{F}^{2}] \leq \delta \mathrm{tr} 
%   \mathbf{V}^{T} \mathbf{V} \leq \frac{d}{\gamma_{2}}.
% \end{equation*}

{\bf Proof of Theorem~\ref{thm:identity}.} Let $\mathbf{P}_n = \mathbf{X}_n \mathbf{X}^{\top}$ and $\mathbf{Q}_n = \mathbf{Y}_n \mathbf{Y}_n^{\top}$. 
For ease of notation, in parts of the proof below we will suppress
  the dependence of ${\bf X}_n$, ${\bf Y}_n$, $\mathbf{P}_n$ and
  $\mathbf{Q}_n$ on $n$ and simply denote these matrices by ${\bf X}$,
  ${\bf Y}$, $\mathbf{P}$, and $\mathbf{Q}$, respectively; we will
  make this dependence explicit when necessary. Suppose that the null
  hypothesis $H_0$ is true, so there exists an orthogonal $\tilde{{\bf
      W}} \in \mathbb{R}^{d \times d}$ such that ${\bf X}={\bf
    Y}\tilde{\bf{W}}$. Let $\alpha$ be given, and let $\eta
  <\alpha/4$. From \eqref{conc_x-xhat}, for all $n$ sufficiently
  large, there exist orthogonal matrices ${\bf W}_X$ and ${\bf W}_Y
  \in \mathcal{O}(d)$ such that with probability at least $1-\eta$,
\begin{align*}
\|\hat{{\bf X}}-{\bf X}{\bf W}_X\|_F &\leq C({\bf X}) + f({\bf X}, \alpha, n)\\
\|\hat{{\bf Y}}-{\bf Y}{\bf W}_Y\|_F &\leq C({\bf Y}) +f({\bf Y}, \alpha, n)
\end{align*}
where $f({\bf X}_n, \alpha, n) \rightarrow 0$ as $n \rightarrow \infty$
for a fixed $\alpha$ and sequence $\{\mathbf{X}_n\}$ satisfying Assumption \ref{eigengap_assump}. 

Let ${\bf W}^*={\bf W}_Y \tilde{{\bf W}} {\bf W}_X$. 
Then there exists
a $n_0 = n_0(\alpha)$ such that for all 
$n>n_0$, with probability at least $1-\eta$, we have
\begin{align*}
  \|\hat{{\bf X}}-\hat{{\bf Y}} {\bf W}^*\|_{F}& % \leq \|\hat{{\bf
  %     X}}-{\bf XW}_X\|_{F} + \|{\bf XW}_X - {\bf Y}\tilde{{\bf W}}{\bf
  %   W}_X\|_{F} +
  % \|{\bf Y}\tilde{{\bf W}}{\bf W}_X - \hat{{\bf Y}} \mathbf{W}^{*}\|_{F} \\
  % &\leq \|\hat{{\bf X}}-{\bf XW}_X\|_{F}+ \|{\bf XW}_X - {\bf
  %   Y}\tilde{{\bf W}} {\bf W}_X\|_{F}+ \|\left({\bf Y}-\hat{{\bf
  %       Y}}{\bf W}_Y\right)
  % \left(\tilde{{\bf W}}{\bf W}_X\right)\|_{F} \\
  \leq \|\hat{{\bf X}}-{\bf XW}_X\|_{F}+
  \|{\bf Y}-\hat{{\bf Y}}{\bf W}_Y\|_{F} \\
  &\leq C({\bf X)}+ C({\bf Y})+f({\bf X}, \alpha, n)+f({\bf Y},
  \alpha, n)
\end{align*}
where we have used the fact that under $H_0$,
$\mathbf{X}=\mathbf{Y}\tilde{\mathbf{ W}}$. We note
that both $C({\bf X})$ and $C({\bf Y})$ are unknown. However, by
Theorem~\ref{thm:conc_Xhat_X}, they can be bounded from above by
$(d\gamma_{2}^{-1}(\mathbf{P}))^{1/2}$ and
$(d\gamma_{2}^{-1}(\mathbf{Q}))^{1/2}$, respectively. Hence for all
$n>n_0$, with probability at least $1 - \alpha$,
% To estimate ${\bf S}_P(d,d)$, we the pooled estimate $[{\bf
% S}_{A^1_n}(d,d)+ {\bf S}_{A^2_n}(d,d)]/2$, and from ProposiBtion
% \eqref{Old_bounds}, we derive that under $H_0$, for all $n>n_0$,
\begin{equation*}
 \frac{\min\limits_{{\bf W} \in \mathcal{O}(d)}\|\hat{{\bf
       X}}_n{\bf W}-\hat{{\bf Y}}_n\|_F} 
{\sqrt{d \gamma_{2}^{-1}(\mathbf{P}_n)} + \sqrt{
    d\gamma_{2}^{-1}(\mathbf{Q}_n)}} \leq 1+r(\alpha,n)
\end{equation*}
where $r(\alpha, n) \rightarrow 0$ as $n\rightarrow \infty$ for a
fixed $\alpha$. In addition, by Proposition~\ref{prop:gamma(A)}, the
terms $\gamma_{2}^{-1}(\mathbf{P}_n)$ and
$\gamma_{2}^{-1}(\mathbf{Q}_n)$ in the denominator can be replaced by
$\gamma_{2}^{-1}(\mathbf{A}_n)$ and $\gamma_{2}^{-1}(\mathbf{B}_n)$
for sufficiently large $n$. Therefore, with probability at least $1 -
\alpha$,
\begin{equation*}
  T_n = \frac{\min\limits_{{\bf W} \in \mathcal{O}(d)}\|\hat{{\bf
       X}}_n{\bf W}-\hat{{\bf Y}}_n\|_F} 
{\sqrt{d \gamma_{2}^{-1}(\mathbf{A}_n)} + \sqrt{
    d\gamma_{2}^{-1}(\mathbf{B}_n)}} \leq 1+\tilde{r}(\alpha,n)
\end{equation*}
where once again, for a fixed $\alpha$, $\tilde{r}(\alpha, n)
\rightarrow 0$ as $n\rightarrow \infty$. We can thus take $n_1 =
n_1(\alpha, C) = \inf \{ n \geq n_0(\alpha) \colon \tilde{r}(\alpha,
n) \leq C - 1\} < \infty$. Then for all $n>n_1$ and $\mathbf{X}_n,
\mathbf{Y}_n$ satisfying $\mathbf{X}_n \upVdash \mathbf{Y}_n$, we
conclude
$$\Pr(T_n \in R) < \alpha.$$

We now prove consistency. Let $$\tilde{\mathbf{W}} = \argmin_{\mathbf{W}
  \in \mathcal{O}(d)} \|\mathbf{X} - \mathbf{Y} \mathbf{W}\|_{F}$$ and
denote by $D(\mathbf{X}, \mathbf{Y}) = \|\mathbf{X} - \mathbf{Y}
\tilde{\mathbf{W}}\|_{F}$.  As before, let ${\bf W}^*={\bf W}_Y
\tilde{{\bf W}} {\bf W}_X$.Note that
\begin{equation*}
  %\label{reverse-triangle}
  \begin{split}
    \|\hat{{\bf X}}-\hat{{\bf Y}}{\bf W}^{*}\|_F &% \geq \|{\bf
    %   Y}\tilde{\bf W}-{\bf X}\|_F - \|\hat{\bf X}-{\bf X}{\bf W}_X +
    % {\bf YW}_Y{\bf
    %   W}-\hat{\bf Y}{\bf W}\|_F \\
    %& 
    \geq D(\mathbf{X}, \mathbf{Y}) - \|\hat{\bf X}-{\bf X}{\bf W}_X
    \|_{F} - \| {\bf YW}_Y -\hat{\bf Y}\|_F
  \end{split}
\end{equation*}
Therefore, for all $n$,
\begin{equation*}
  \begin{split}
    \mathbb{P}(T_n \not \in R) &\leq \mathbb{P}\biggl(
    \frac{\|\hat{{\bf X}}-\hat{{\bf Y}}{\bf W}^{*}\|_F}{ \sqrt{d
        \gamma_{2}^{-1}(\mathbf{A})} + \sqrt{
        d\gamma_{2}^{-1}(\mathbf{B})}} \leq C\biggr) \\
    % & \leq \mathbb{P}\biggl(\frac{\|{\bf Y} \tilde{\bf W} -{\bf X}\|_F
    %   - \|\hat{\bf X}-{\bf X}{\bf W}_X \|_{F} - \| {\bf YW}_Y
    %   -\hat{\bf Y}\|_F}{\sqrt{d \gamma_{2}^{-1}(\mathbf{A})} + \sqrt{
    %     d\gamma_{2}^{-1}(\mathbf{B})}} \leq C \biggr) \\
    &= \mathbb{P}\Bigl( \|\hat{\bf X}-{\bf X}{\bf W}_X \|_{F} + \|
    {\bf YW}_Y -\hat{\bf Y}\|_F + C' \geq D(\mathbf{X}, \mathbf{Y})
    \Bigr) % \\
% & \leq \mathbb{P}(\|\hat{\bf X}-{\bf X}{\bf W}_X \|_{F} \geq
% D(\mathbf{X}, \mathbf{Y}) -  C') + \mathbb{P}(\|\hat{\bf Y}-{\bf
%   X}{\bf W}_Y \|_{F} \geq D(\mathbf{X}, \mathbf{Y})  -  C') 
  \end{split}
\end{equation*}
where $C' = C(\sqrt{d \gamma_{2}^{-1}(\mathbf{A})} + \sqrt{
  d\gamma_{2}^{-1}(\mathbf{B})})$.
% We observe that the concentration bound in
% Theorem~\ref{thm:conc_Xhat_X} guarantees that there exists some
% constant $M > 0$ such that % for any there exists a
% % constant $D(X,Y)$ such that for $\epsilon>0$, there exists $n_2$ such
% % that for all $n>n_2$,
% \begin{equation*}
%  \|\hat{\bf X}-{\bf X}{\bf W}_X + {\bf YW}_Y{\bf W}-\hat{\bf Y}{\bf W}\|_F 
% - C(\mathbf{X}) - C(\mathbf{Y}) < M \quad \text{almost surely.}
% \end{equation*}
% See also the statement of
% Corollary~\ref{cor:1}. Now $C({\bf X})$ is bounded from above by $(d
% \gamma_{2}^{-1}(\mathbf{X} \mathbf{X}^{T}))^{1/2}$ and similarly for $C({\bf
%   Y})$. 
By Assumption~\ref{eigengap_assump}, there exists some $n_0$ and some
$c_0 > 0$ such that $\gamma_{2}(\mathbf{P}_n) \geq c_0$ and
$\gamma_{2}(\mathbf{Q}_n) \geq c_0$ for all $n \geq
n_0$. % Hence, for all $n \geq n_0(\alpha)$, $C'$
% is bounded from above by some constant $M_0 > 0$.  _
% \begin{equation*} 
%    \|\hat{\bf X}-{\bf X}{\bf W}_X + {\bf YW}_Y{\bf W}-\hat{\bf Y}{\bf W}\|_F 
% - 2\sqrt{d/c_0} < M \quad \text{almost surely}.
% \end{equation*}
% Therefore, 
Now, let $\beta > 0$ be given. By the almost sure convergence of
$\|\hat{\mathbf{X}}-\mathbf{XW}_x\|_F$ to $C({\bf X})$, established in
Theorem \ref{thm:conc_Xhat_X}, and the almost sure convergence of
$\gamma_2(\mathbf{A})$ to $\gamma_2(\mathbf{P})$ given in
\ref{prop:gamma(A)}, we deduce that there exists a constant
$M_1(\beta)$ and a positive integer $n_0 = n_0(\alpha, \beta)$ so
that, for all $n \geq n_0(\alpha, \beta)$,
\begin{gather*}
  \mathbb{P}(\|\hat{\bf X}-{\bf X}{\bf W}_X \|_{F} + C\sqrt{d
    \gamma_{2}^{-1}(\mathbf{A})} \geq M_1/2) \leq \beta/2 \\  
  \mathbb{P}(\|\hat{\bf Y}-{\bf Y}{\bf W}_Y \|_{F} + C \sqrt{d
    \gamma_{2}^{-1}(\mathbf{B})} 
\geq M_1/2) \leq \beta/2
\end{gather*}
%By Theorem~\ref{thm:conc_Xhat_X} and Proposition~\ref{prop:gamma(A)}, $M_1$
%nd $n_0(\alpha, \beta)$ exists for all choice of $\beta$. 
If $b_n\rightarrow \infty$, there exists some $n_2 = n_2(\alpha, \beta, C)$ such
that, for all $n \geq n_2$, either $D(\mathbf{X}_n, \mathbf{Y}_n) =0$ or $D(\mathbf{X}_n, \mathbf{Y}_n) \geq
M_1$. Hence, for all $n \geq n_2$, if $D(\mathbf{X}_n, \mathbf{Y}_n) \neq 0$, then $\mathbb{P}(T_n \not \in R) \leq
\beta$, i.e., our test statistic $T_n$ lies within the rejection
region $R$ with probability at least $1 - \beta$, as required.
\qed

{\bf Proof of Theorem~\ref{thm:2}.}
  The proof of this result is almost identical to that of
  Theorem~\ref{thm:identity}. We sketch here the necessary
  modifications. As before, we suppress dependence on $n$ unless
  necessary. Let $\alpha$ be given and let $\eta = \alpha/4$. By
  Theorem~\ref{thm:conc_Xhat_X}, for $n$ sufficiently large, there
  exists some orthogonal $\mathbf{W}_{\mathbf{X}} \in \mathcal{O}(d)$
  such that, with probability at least $1 - \eta$
  \begin{equation*}
    \|\hX \mathbf{W}_{\mathbf{X}} - \mathbf{X} \|_{F} \leq C(\mathbf{X}) +
    f(\mathbf{X}, \alpha, n)
  \end{equation*}
  where for any fixed $\alpha$, $f(\mathbf{X}_n,
  \alpha, n) \rightarrow 0$ as $n \rightarrow \infty$ and
  $\{\mathbf{X}_n\}$ satisfies Assumption~\ref{eigengap_assump}. % From the form
%   of $C(\mathbf{X})$ given in Theorem \ref{thm:conc_Xhat_X}, for all
%   $n$,
% \begin{equation}\label{eq:C(X)-bounds}
% C(\mathbf{X}_n) \leq \sqrt{\frac{d}{\gamma_2(\mathbf{P}_n)}}
% \end{equation}
Now, again for $n$ sufficiently large,
  \begin{equation*}
    \begin{split}
     \|\hX \mathbf{W}_{\mathbf{X}}/\|\hX\|_{F} -
    \mathbf{X}/\|\mathbf{X}\|_{F}\|_{F} &\leq \frac{\|\hX \mathbf{W}_{\mathbf{X}}
    - \mathbf{X}\|_{F}}{\|\hX\|_{F}} + \|\mathbf{X}\|_{F} \Bigl|\frac{1}{\|\hX\|_{F}}
    - \frac{1}{\|\mathbf{X}\|_{F}}\Bigr| \\ &\leq \frac{\|\hX \mathbf{W}_{\mathbf{X}}
    - \mathbf{X}\|_{F}}{\|\hX\|_{F}} + \frac{|\|\hX\|_{F} -
    \|\mathbf{X}\|_{F}|}{\|\hX\|_{F}} \\
   & \leq \frac{\|\hX \mathbf{W}_{\mathbf{X}}
    - \mathbf{X}\|_{F}}{\|\hX\|_{F}} + \frac{|\|\hX \mathbf{W}_{\mathbf{X}}\|_{F} -
    \|\mathbf{X}\|_{F}|}{\|\hX\|_{F}} \\
   & \leq \frac{2 \|\hX \mathbf{W}_{\mathbf{X}}
    - \mathbf{X}\|_{F}}{\|\hX\|_{F}} \leq \frac{2( C(\mathbf{X}) +
    f(\mathbf{X}, \alpha, n))}{\|\hX\|_{F}}
    \end{split}
  \end{equation*}
  with probability at least $1 - \eta$. An analogous bound can also be
  derived for $\mathbf{Y}$. Under the null
  hypothesis, $\mathbf{X} \upVdash c \mathbf{Y}$ for some 
  $c > 0$, so we derive that
  \begin{equation*}
    \min\limits_{{\bf W} \in \mathcal{O}(d)} 
\|\hat{{\bf X}}{\bf W}/\|\hX\|_{F} - \hat{{\bf
      Y}}/\|\hat{\mathbf{Y}}\|_{F} \|_{F} \leq \frac{2(C(\mathbf{X}) +
    f(\mathbf{X}, \alpha, n))}{\|\hX\|_{F}} + \frac{2(C(\mathbf{Y}) +
    f(\mathbf{Y}, \alpha, n))}{\|\hat{\mathbf{Y}}\|_{F}}.
  \end{equation*}
  We thus conclude that for $n$ sufficiently large,
  \begin{equation*}
 T_n=\frac{\min\limits_{{\bf W} \in \mathcal{O}(d)} 
\|\hat{{\bf X}}{\bf W}/\|\hX\|_{F} - \hat{{\bf
      Y}}/\|\hat{\mathbf{Y}}\|_{F} \|_{F}}
{2 \sqrt{d \gamma^{-1}_2(\mathbf{A})}/\|\hX\|_{F}+ 2 \sqrt{d
    \gamma^{-1}_2(\mathbf{B})}/\|\hat{\mathbf{Y}}\|_{F}} \leq 1 + r(\alpha, n)
\end{equation*}
where $r(\alpha, n) \rightarrow 0$ as $n \rightarrow \infty$ for a
fixed $\alpha$. We can now choose a $n_1 = n_1(\alpha, C)$ for
which $r(\alpha, n_1) \leq C - 1$. This implies that for all $n \geq n_1$,
$\mathbb{P}(T_n \in R) \leq \alpha$
which establishes that the test statistic $T_n$ with rejection region $R$ is an
at most level-$\alpha$ test.  The proof of consistency proceeds in an
almost identical manner to that in Theorem~\ref{thm:identity} and we omit the details. \qed

{\bf Proof of Theorem~\ref{thm:1}}
We first show that the test statistic as defined along with the
rejection region $R = \{T > 1\}$ is asymptotically an at-most
level-$\alpha$ test. We have, for any $\mathbf{W} \in \mathcal{O}(d)$,  
  \begin{equation*}
    \begin{split}
    \|\mathcal{P}(\hX) \mathbf{W} - \mathcal{P}(\mathbf{X}) \|_{F} & = \|
    \mathcal{D}^{-1}(\hX) \hX \mathbf{W} - \mathcal{D}^{-1}(\hX) \mathbf{X} +
    \mathcal{D}^{-1}(\hX) \mathbf{X} - \mathcal{D}^{-1}(\mathbf{X}) \mathbf{X}
    \|_{F} \\
    & \leq \|\mathcal{D}^{-1}(\hX)\|_{2} \|\hX \mathbf{W} - \mathbf{X} \|_{F} +
    \|(\mathcal{D}^{-1}(\hX) - \mathcal{D}^{-1}(\mathbf{X})) \mathbf{X}\|_{F} % \\
    % & \|(\mathcal{D}(\hX))^{-1}\|_{F} \|\hX - \mathbf{X} \| +
    % \|\mathbf{X} \| \| (\mathcal{D}(\hX)) -
    % (\mathcal{D}(\mathbf{X})) \|_{F} \| (\mathcal{D}(\hX))^{-1}
    % (\mathcal{D}(\hX))^{-1} \|
    \end{split}
  \end{equation*}
  The term $  \|(\mathcal{D}^{-1}(\hX) -
  \mathcal{D}^{-1}(\mathbf{X})) \mathbf{X}\|_{F}$ can be
  written as
  \begin{equation*}
    \begin{split}
      \|(\mathcal{D}^{-1}(\hX) -
  \mathcal{D}^{-1}(\mathbf{X})) \mathbf{X}\|_{F}^{2} &=
  \sum_{i=1}^{n} \|X_i \|^{2} \Bigl(\frac{1}{\|\hat{X}_i\|} -
      \frac{1}{\|X_i\|} \Bigr)^{2} 
       = \sum_{i=1}^{n} \frac{(\|\mathbf{W} X_i\| -
         \|\hat{X}_i\|)^2}{\|\hat{X}_i\|^2}  % \leq \sum_{i=1}^{n} \frac{\|\mathbf{W} X_i -
         % \hat{X}_i\|^2}{\|\hat{X}_i\|^2} 
       \leq \frac{\|\hX \mathbf{W} - \mathbf{X} \|^{2}_{F}}{ \min_{i} \|\hat{X}_i\|_{2}^{2}} 
    \end{split}
  \end{equation*}
  and hence,
  \begin{equation}
    \label{eq:projection-bound}
    \|\mathcal{P}(\hX) \mathbf{W} -
    \mathcal{P}(\mathbf{X}) \|_{F} \leq 2 \|\hX \mathbf{W} -
    \mathbf{X} \|_{F} \|\mathcal{D}^{-1}(\hX) \|_{2}
  \end{equation}
  % Now, let $\alpha \in (0,1)$ be given and set $\eta = \alpha/4$. Then 
  % by Lemma~\ref{lem:Xhat_X_2_to_infty}, for $n$ sufficiently large, there
  % exists some orthogonal $\mathbf{W}_{\mathbf{X}}$ such that, with
  % probability at least $1 - \eta$, 
  %  \begin{equation}
  %    \label{eq:13}
  %    \|\mathcal{P}(\hX) \mathbf{W}_{\mathbf{X}} -
  %    \mathcal{P}(\mathbf{X}) \|_{F} \leq \frac{170 d^{3/2}
  %      \log{(n/\eta)}}{ \sqrt{\gamma_{1}^{7}(\mathbf{P}) \delta(\mathbf{P})}} 
  %    \|\mathcal{D}^{-1}(\hX)\|_{F} = S(\mathbf{P}) \|\mathcal{D}^{-1}(\hX)\|_{F}.
  %  \end{equation}
   An analogous bound holds for $    \|\mathcal{P}(\hat{\mathbf{Y}})
   \mathbf{W} -
    \mathcal{P}(\mathbf{Y}) \|_{F}$. % Now, let $\alpha \in (0,1)$ be given and set $\eta = \alpha/4$. 
   % Under the null hypothesis $\mathbf{X}
   % \upVdash \mathbf{D} \mathbf{Y}$, we have
    Therefore, 
   % \begin{equation}
   %   \label{eq:11}
   %   \min_{\mathbf{W} \in \mathcal{O}(d)} \|\mathcal{P}(\hX) \mathbf{W} -
   %   \mathcal{P}(\hat{\mathbf{Y}})\|_{F} \leq S(\mathbf{P})
   %   \|\mathcal{D}^{-1}(\hX) \|_{F} + S(\mathbf{Q})
   %   \|\mathcal{D}^{-1}(\hat{\mathbf{Y}}) \|_{F}
   % \end{equation}
   % and hence,
   \begin{equation*}
      \frac{\min_{\mathbf{W} \in \mathcal{O}(d)} \|\mathcal{P}(\hX) \mathbf{W} -
     \mathcal{P}(\hat{\mathbf{Y}})\|_{F}}{2 \|\hX \mathbf{W}_{\mathbf{X}} - \mathbf{X}\|_{F}
     \|\mathcal{D}^{-1}(\hX) \|_{2} + 2 \|\hat{\mathbf{Y}} \mathbf{W}_{\mathbf{Y}} - \mathbf{Y}\|_{F}
     \|\mathcal{D}^{-1}(\hat{\mathbf{Y}}) \|_{2}} \leq 1
   \end{equation*}
   We can now replace $\|\hX
   \mathbf{W}_{\mathbf{X}} - \mathbf{X}\|_{F}$ by $\sqrt{d
     \gamma_{2}^{-1} (\mathbf{A})}$ and $\|\hat{\mathbf{Y}}
   \mathbf{W}_{\mathbf{Y}} - \mathbf{Y}\|_{F}$ by $\sqrt{d \gamma_{2}^{-1}
     (\mathbf{B})}$ to yield
  \begin{equation*}
     T_n =   \frac{\min_{\mathbf{W} \in \mathcal{O}(d)} \|\mathcal{P}(\hX) \mathbf{W} -
     \mathcal{P}(\hat{\mathbf{Y}})\|_{F}}{2 \sqrt{d
     \gamma_{2}^{-1}(\mathbf{A})} \|\mathcal{D}^{-1}(\hX) \|_{2}  +
   2 \sqrt{d \gamma_{2}^{-1} (\mathbf{B})}
   \|\mathcal{D}^{-1}(\hat{\mathbf{Y}}) \|_{2}} 
 \leq 1 + r(\alpha, n)
   \end{equation*}
   where $r(\alpha, n) \rightarrow 0$ as $n \rightarrow\infty$ for a
   fixed $\alpha$. We can therefore choose a $n_1 = n_1(\alpha, C)$
   for which $r(\alpha, n_1) \leq C - 1$. This implies that
   for all $n \geq n_1$, $\mathbb{P}(T_n \in C) \leq \alpha$
   yielding that the test statistic $T_n$ with rejection region $R$ is
   an at most level-$\alpha$ test. 
   
   We now prove consistency of this test procedure.  Suppose 
   the sequences of latent positions $\{\mathbf{X}_n\}$ and $\{\mathbf{Y}_n\}$
   are such that $\mathbf{X}_n \nupVdash \mathbf{D}_n \mathbf{Y}_n$. Denote by
   $h(\hat{\mathbf{X}}, \hat{\mathbf{Y}})$ and $f(\hat{\mathbf{X}}, \hat{\mathbf{Y}})$ the ratios
   \begin{gather*}
     h(\hat{\mathbf{X}}, \hat{\mathbf{Y}}) = 
\frac{\sqrt{
     \gamma_{2}^{-1}(\mathbf{A})} \|\mathcal{D}^{-1}(\hX) \|_{2}  +
   \sqrt{\gamma_{2}^{-1} (\mathbf{B})}
   \|\mathcal{D}^{-1}(\hat{\mathbf{Y}}) \|_{2}}{\sqrt{
     \gamma_{2}^{-1}(\mathbf{P})} \|\mathcal{D}^{-1}(\mathbf{X}) \|_{2}  +
   \sqrt{ \gamma_{2}^{-1} (\mathbf{Q})}
   \|\mathcal{D}^{-1}(\mathbf{Y}) \|_{2}} \\
   f(\hat{\mathbf{X}}, \hat{\mathbf{Y}}) =
     \frac{\|\mathcal{P}(\hat{\mathbf{X}}) \mathbf{W}_{\mathbf{X}}
       - \mathcal{P}(\mathbf{X}) \|_{F} +
       \|\mathcal{P}(\hat{\mathbf{Y}}) \mathbf{W}_{\mathbf{Y}}
       - \mathcal{P}(\mathbf{Y}) \|_{F}}{2 \sqrt{d
     \gamma_{2}^{-1}(\mathbf{A})} \|\mathcal{D}^{-1}(\hX) \|_{2}  + 2
   \sqrt{d \gamma_{2}^{-1} (\mathbf{B})}
   \|\mathcal{D}^{-1}(\hat{\mathbf{Y}}) \|_{2}}
   \end{gather*}
   Then, for all $n$, 
   \begin{equation*}
     \begin{split}
     \mathbb{P}(T_n \not \in R_n) &\leq \mathbb{P}\biggl(
      \frac{ \min_{W \in \mathcal{O}(d)}\|\mathcal{P}(\mathbf{X}) \mathbf{W} -
        \mathcal{P}(\mathbf{Y}) \|_{F} }{2 \sqrt{d
     \gamma_{2}^{-1}(\mathbf{A})} \|\mathcal{D}^{-1}(\hX) \|_{2}  + 2
   \sqrt{d \gamma_{2}^{-1} (\mathbf{B})}
   \|\mathcal{D}^{-1}(\hat{\mathbf{Y}}) \|_{2}}
 \leq C + f(\hat{\mathbf{X}}, \hat{\mathbf{Y}})\biggr)
      \\ & \leq
      \mathbb{P}\biggl( h(\hat{\mathbf{X}}, \hat{\mathbf{Y}})(C +
      f(\hat{\mathbf{X}}, \hat{\mathbf{Y}})) \geq 
\frac{ \min\limits_{{\bf W} \in \mathcal{O}(d)}\| \mathcal{P}({\bf
    X}_n) {\bf W} -
 \mathcal{P}({\bf Y}_n)
 \|_{F}}{2 \sqrt{d \gamma_{2}^{-1}(\mathbf{P})}
   \|\mathcal{D}^{-1}(\mathbf{X})\|_{2} + 2 \sqrt{d \gamma_{2}^{-1}(\mathbf{Q})} 
 \|\mathcal{D}^{-1}(\mathbf{Y})\|_{2}} \biggr)
\end{split}
  \end{equation*}
 
   Now, for a given $\beta > 0$, let $M_1 = M_1(\beta)$ and $n_0 =
   n_0(\alpha, \beta)$ be such that, for all $n \geq n_0(\alpha,
   \beta)$, 
   \begin{equation*}
     \mathbb{P}(C + f(\hat{\mathbf{X}}, \hat{\mathbf{Y}}) \geq M_1)
     \leq \beta/2.
   \end{equation*}
   By Eq.~\eqref{eq:projection-bound} and Proposition~\ref{prop:gamma(A)}, $M_1(\beta)$ and
   $n_0(\alpha,\beta)$ exists for all choice of $\beta$. 
   We now show
   that there exists, for any $\beta > 0$, some $n_1 = n_1(\beta)$ such that, for all $n \geq
   n_1(\beta)$, 
   \begin{equation}
     \label{eq:14}
     \mathbb{P}(h(\hat{\mathbf{X}}, \hat{\mathbf{Y}}) \geq 4) \leq \beta/2.
   \end{equation}
   Indeed, 
   \begin{equation*}
 h(\hat{\mathbf{X}}, \hat{\mathbf{Y}}) \leq \max\biggl\{
\frac{\sqrt{
     \gamma_{2}^{-1}(\mathbf{A})} \|\mathcal{D}^{-1}(\hX) \|_{2}}{\sqrt{
     \gamma_{2}^{-1}(\mathbf{P})} \|\mathcal{D}^{-1}(\mathbf{X})
   \|_{2}}, \frac{
   \sqrt{\gamma_{2}^{-1} (\mathbf{B})}
   \|\mathcal{D}^{-1}(\hat{\mathbf{Y}}) \|_{2}}{
   \sqrt{ \gamma_{2}^{-1} (\mathbf{Q})}
   \|\mathcal{D}^{-1}(\mathbf{Y}) \|_{2}}\biggr\} 
% \frac{S(\mathbf{P})
%        \|\mathcal{D}^{-1}(\mathbf{X})\|_{F}}{S(\mathbf{A})
%         \|\mathcal{D}^{-1}(\hat{\mathbf{X}}) \|_{F}}, \frac{S(\mathbf{Q})
%        \|\mathcal{D}^{-1}(\mathbf{Y})\|_{F}} {S(\mathbf{B})
%        \|\mathcal{D}^{-1}(\hat{\mathbf{Y}})\|_{F}} \Bigr\}
   \end{equation*}
   In addition, we have
   \begin{equation*}
     \begin{split}
     \frac{\|\mathcal{D}^{-1}(\mathbf{X})\|_{2}}{\|\mathcal{D}^{-1}(\hat{\mathbf{X}})\|_{2}}
     &= \frac{1/\min_{i} \|X_i\|_{2}}{1/\min_{i} \|\hat{X}_i\|_2} =
     \frac{\min_{i} \|\hat{X}_i \|_{2}}{\min_{i} \|X_i \|_{2}} \leq  
     \max_{i}
        \frac{\|\hat{X}_i\|}{\|X_i\|} \leq 1 +
        \frac{\max_{i} \|\mathbf{W} \hat{X}_i - X_i\|_{2}}{\min_{i} \|X_i\|_{2}}
     \end{split}
   \end{equation*}
   for any orthogonal matrix $\mathbf{W}$. We now use the following
   result, namely Lemma 5 from \citet{lyzinski13:_perfec}, to bound
   the maximum of the $l_2$, norm of the rows of $\hX \mathbf{W} - \mathbf{X}$.
   % While this bound is sufficient for our current purpose, it is
   % somewhat loose and we believe a tighter bound will lead to tighter control of the resulting test
   % statistic.
   \begin{lemma}\label{lem:Xhat_X_2_to_infty}
  % Assume the conditions in Assumption~\ref{eigengap_assump} are
  % satisfied.  
     Suppose Assumption \ref{eigengap_assump_diagonal} holds, and let $c
     > 0$ be arbitrary. Then there exists a $n_0 (c)$ such that for all
     $n>n_0$ and $n^{-c} <\eta<1/2$, there exists a deterministic ${\bf
       W}={\bf W}_n \in \mathcal{O}(d)$ such that, with probability at
     least $1 - 3\eta$,
     \begin{equation}\label{conc_x-xhat2toinfty}
       \max_{i}
       \| \hat{X}_i - \mathbf{W} X_i \|  \leq \frac{85 d^{3/2}
         \log{(n/\eta)}}{\sqrt{ \gamma_{1}^{7}(\mathbf{P}) \delta(\mathbf{P})}}.
     \end{equation}
   \end{lemma}
   Continuing with the proof of the theorem, by Lemma~\ref{lem:Xhat_X_2_to_infty} and the conditions in
   Assumption~\ref{eigengap_assump_diagonal} on $\min_{i} \|X_i\|$, there exist some
   $n_1(\beta)$ such that for all $n \geq n_1(\beta)$, 
   \begin{equation*}
         \mathbb{P}\biggl(  1 +
        \frac{\max_{i} \|\mathbf{W} \hat{X}_i - X_i\|_{2}}{\min_{i} \|X_i\|_{2}}
\geq 2 \biggr) \leq \beta/8.
   \end{equation*}
   Proposition~\ref{prop:gamma(A)} then implies that there exist
   some $n_1(\beta)$ such that for all $n \geq n_1(\beta)$, 
    \begin{equation*}
      \mathbb{P}\biggl(
\frac{\sqrt{
     \gamma_{2}^{-1}(\mathbf{A})} \|\mathcal{D}^{-1}(\hX) \|_{2}}{\sqrt{
     \gamma_{2}^{-1}(\mathbf{P})} \|\mathcal{D}^{-1}(\mathbf{X})
   \|_{2}} \geq 4\biggr) \leq \beta/4.
    \end{equation*}
   The same argument can be applied to the ratio depending on
   $\hat{\mathbf{Y}}$ and $\mathbf{Y}$. % Thus
   % there exists some $n_1(\beta)$ such that Eq.~\eqref{eq:14} holds
   % for all $n \geq n_1(\beta)$.
   Since
   $D_{\mathcal{P}}(\mathbf{X}_n, \mathbf{Y}_n) \rightarrow \infty$, there
   exists some $n_2 = n_2(\alpha, \beta, C)$ such that for all $n \geq
   n_2$, 
   \begin{equation*}
     \frac{ \min\limits_{{\bf W} \in \mathcal{O}(d)}\| \mathcal{P}({\bf
    X}_n) {\bf W} -
 \mathcal{P}({\bf Y}_n)
 \|_{F}}{2 \sqrt{d \gamma_{2}^{-1}(\mathbf{P})}
   \|\mathcal{D}^{-1}(\mathbf{X})\|_{2} + 2 \sqrt{d \gamma_{2}^{-1}(\mathbf{Q})} 
 \|\mathcal{D}^{-1}(\mathbf{Y})\|_{2}}  \geq 4 M_1
   \end{equation*}
   Hence for all $n \geq n_2$, $\mathbb{P}(T_n \not \in
   R) \leq \beta$ as required.  \qed
\end{document}